\documentclass[twocolumn,prd,superscriptaddress,amssymb,nofootinbib,amsmath,nobibnotes,aps,prlgroupedaddress]{revtex4-2}  % for review and submission
\usepackage{graphicx}  % needed for figures
\usepackage{dcolumn}   % needed for some tables
\usepackage{bm}        % for math
\usepackage{amssymb}   % for math
\usepackage[]{amsmath,amssymb}
\usepackage{graphics,epsfig}
\usepackage{float}
\usepackage{mathtools}
\usepackage{esint}
\usepackage[utf8]{inputenc}
\usepackage[english]{babel}
\usepackage{mathtools}
\usepackage{xcolor}
\usepackage{float}  
\usepackage{setspace} 
\usepackage{indentfirst}  
\usepackage{cancel}  
\usepackage{soul}
\usepackage{hyperref}  
\hypersetup{
            colorlinks=true,
            linkcolor=black,
            urlcolor=blue,
            } 
\usepackage{tcolorbox}   
\usepackage{orcidlink}

\newcommand{\mchirp}{\mathcal{M}}

\begin{document}

%% Title
% \title{Gravitational waveform of compact binaries with an ultralight vector dark matter environment}
\title{Gravitational waves of quasi-circular, inspiraling black hole binaries \\ in  an ultralight vector dark-matter environment}
%% %simple case: 2 authors, same institution
\author{Tomás Ferreira Chase \orcidlink{0009-0001-0286-2136}} 
\email{tferreirachase@df.uba.ar}
\affiliation{Universidad de Buenos Aires, Facultad de Ciencias Exactas y Naturales, \\ Departamento de Física, Buenos Aires, Argentina.} 
\affiliation{CONICET - Universidad de Buenos Aires, \\ Instituto de Física de Buenos Aires (IFIBA), Buenos Aires, Argentina.}
\author{Diana López Nacir \orcidlink{0000-0003-4398-1147}} 
\email{dnacir@df.uba.ar}
\affiliation{Universidad de Buenos Aires, Facultad de Ciencias Exactas y Naturales, \\ Departamento de Física, Buenos Aires, Argentina.} 
\affiliation{CONICET - Universidad de Buenos Aires, \\ Instituto de Física de Buenos Aires (IFIBA), Buenos Aires, Argentina.}
\author{Nicolás Yunes \orcidlink{0000-0001-6147-1736}} 
\email{nyunes@illinois.edu}
\affiliation{Illinois Center for Advanced Studies of the Universe, Department of Physics, University of Illinois at Urbana-Champaign, Urbana, Illinois 61801, USA.}

\date{\today}

\begin{abstract}
    The gravitational waves emitted by massive black hole binaries can be affected by a variety of environmental effects, which, if detected, could inform astrophysics and cosmology.  
    We here study how gravitational waves emitted by black holes in quasi-circular orbits are affected by the presence of an ultra-light, vector-field, dark-matter environment that is minimally coupled to the binary. 
    This dark-matter environment induces oscillatory gravitational potentials that perturb the orbit of the binary, leaving an imprint in the binary's binding energy, and thus, on the gravitational waves emitted. 
    We here compute the effect of this environment on the gravitational-wave phase using the stationary-phase approximation within the post-Newtonian formalism. 
    We then perform a Fisher analysis to estimate the detectability of this environmental effect with a four-year LISA observation, focusing on vector fields with ultra-light masses in the $(10^{-19}, 10^{-16}) \; \rm{eV}$ range.
    We conclude that the observation of such gravitational waves with space-borne interferometers, like LISA, could yield a measurement or constraint on local, vector dark-matter environments, provided the dark-matter density is larger than roughly $10^{14} \rm{M}_\odot/{\rm{pc}}^3$.
%    We conclude that the effect is negligible in the signal unless the dark-matter density surrounding the black holes is several orders of magnitude higher than that in the Solar System.
\end{abstract}
\maketitle

%%%%%%%%%%%%%%%%%%%%%%%%%%%%%%%%%%%%%%%%%%%%%%%%%%%%%%%%%%%%%%%%%%%%%%%%%%%%%%%%
\section{Introduction}

Although there is strong evidence that dark matter represents approximately $25\%$ of the energy content of the universe today, its nature still remains unknown. Many dark-matter models (be it wave-like or particle-like) have been considered, with a wide range of masses and interactions. The lightest class of candidates is usually referred to as \emph{ultralight dark-matter} field models, where the mass of the field that describes the dark matter ranges from $(10^{-28},1) \; \rm{eV}$. These candidates have gained attention in recent years due to their agreement with the fiducial $\Lambda$CDM model at large scales and their different predictions at small scales \cite{Ferreira:2020fam, Hlozek:2014lca, Chase:2023puj, Chase:2024wsq}.

One distinctive feature of ultralight dark-matter candidates is that they induce oscillatory gravitational potentials. The frequency of the oscillations is determined essentially by the mass of the ultralight dark-matter particles. In cosmological contexts, the dark-matter field  oscillations usually average out, but this need not be the case around astrophysical systems. In fact, previous work has shown that an oscillatory ultralight, dark-matter environment induces oscillatory behavior in the spacetime metric of astrophysical systems, for example impacting the orbital dynamics and evolution of binary pulsars \cite{Blas:2019hxz, Blas:2016ddr, LopezNacir:2018epg}. Could such an ultralight, dark-matter environment also affect the gravitational waves emitted by compact binaries? 

%{\ny{Tomi, it can be considered ``bad form'' to relabel the labels of the bib references, because when you do so, it's hard to find them in Inspires-Hep. The latter actually has a bib generator that is quite handy, so you should try to stick to that in the future. }}

General relativity predicts that gravitational waves are excited by any system with a non-trivial, time-varying multipole-moment structure, such as due to the orbital acceleration of a binary system. When in isolation, the orbital dynamics and evolution of inspiraling compact binaries can be determined from perturbative solutions to the Einstein equations. In practice, when the compact binary is sufficiently well-separated, one employs a post-Minkowskian expansion (in weak fields) or a post-Newtonian expansion (in both weak fields and small velocities) to calculate the gravitational waves in the inspiral phase of coalescence from first principles~\cite{Blanchet:2013haa}. When not in isolation, for example due to the presence of an accretion disk~\cite{Yunes:2011ws, Kocsis:2011dr, Speri:2022upm} or a dark-matter environment~\cite{Cole:2022yzw,Barausse:2014tra}, the orbital dynamics and evolution of inspiraling compact binaries is in general modified, leaving an environmental imprint on the gravitational waves generated. The question \textit{is} then not whether an ultralight, dark-matter environmental imprint is present in gravitational waves, but rather whether this imprint is \textit{detectable}.        

%, and the waveforms can be computed analytically with high precision for several sources, such as for binary systems. Since these predictions are made from first principles, the detection of gravitational wave provides some of the most stringent tests to GR.

% The answer to this question depends both on the properties of the ULDM environment considered, as well as on specifics of the binary system. Ultra-light scalar clouds are expected to grow around supermassive black holes through superradiance, thus draining the spin angular momentum of the parent black hole. In turn, these clouds can both emit gravitational waves themselves due to energy transitions~\cite{}, as well as affect the dynamics of smaller or comparable compact objects in orbit around the parent black hole. All of these effects are expected to leave strong signatures in the gravitational waves emitted by such systems, if an ultra-light scalar cloud is present~\cite{}.  

One of the most studied phenomena connecting dark-matter fields and black holes is the \textit{superradiant instability}, which leads to the formation of a bosonic condensate around spinning black holes \cite{Baumann:2018vus, Boskovic:2024fga, Baryakhtar:2017ngi, Pani:2012vp}. Such studies lead to the following exclusion windows for the mass of the vector field  at 68\% confidence limit \cite{Stott:2020gjj}: {$6.2\times10^{-15}  {\rm eV} \leq m \leq 3.9\times 10^{-11}  {\rm eV}, \, 2.8\times 10^{-22} {\rm eV} \leq m \leq 1.9\times10^{-16} {\rm eV}$}, which incorporate observations of the event GW190521 and the shadow of M87*.  In this paper we focus on a  different phenomena than the one studied in previous work. In our set-up, for the sake of simplicity, we consider black holes with negligible spins so that  the formation of a condensate via supperradiance can be neglected. 
%We also argue that the backreaction of the binary system over the dark-matter field can be neglected at the astrophysical scales of relevance to our work (see Appendix \ref{appendix:backreaction}).

%As we discuss in Sec.~\ref{sec:impact_vector-field dark matter_orbit}, for the field masses and black holes considered here, the dark-matter field is approximately homogeneous at the characteristic scale of the binary systems. Furthermore, we neglect the spin of the black holes. Then, for the sake of simplicity, we neglect the formation of the condensate in our set-up. We also argue that the backreaction of the binary system over the dark-matter field can be neglected at the astrophysical scales of relevance to our work (see Appendix \ref{appendix:backreaction}).

Recent efforts have also been made to search for the direct imprint of ultralight dark matter in gravitational waves. For example, in \cite{Brax:2024yqh, Marriott-Best:2025sez}, the dephasing of gravitational waves produced by dark-matter fields was considered. In these works, the dephasing is produced in the propagation of the gravitational-wave signal from the source to the detector, caused by the gravitational potentials generated by the ultralight dark-matter fields. In this work, however, we focus on how these gravitational potentials create a dephasing in the \textit{generation} of the gravitational-wave signal, by changing the world-line of the binary system components, rather than by correcting the way the gravitational waves propagate.

When the binary systems emitting gravitational waves  is immersed in a virialized ultralight, dark-matter halo, the dark matter that forms the halo (be it scalar or vector particles) can be described as a collection of waves with random phases~\cite{Amin:2022pzv, Centers:2019dyn, Schive:2014dra, Lopez-Sanchez:2025osk}. These waves induce oscillations in the spacetime metric, which then perturb the orbital motion of the binary. In particular, the ultralight dark-matter oscillations induce a correction to the acceleration of the binary components, which then affects the orbital energy and angular momentum, introducing dark-matter corrections to Kepler's third law. The ultralight dark-matter environment, therefore, perturbs the orbital motion of the binary, which then leaves an imprint on the time-varying multipole-moment structure of the spacetime, and thus, on the gravitational waves emitted. Given this, one expects that the strength of the dark-matter imprint on the emitted gravitational waves to be proportional to the dark-matter density and sensitive to the dark-matter mass, which sets the dark-matter oscillation scale, specially close to resonances between this oscillation scale and the orbital timescale. 

\begin{figure}[th]
    % \centering
    %\includegraphics[width=\linewidth]{Figures/Fig_intro_Nico.pdf}
    %\\ 
    \includegraphics[width=\linewidth]{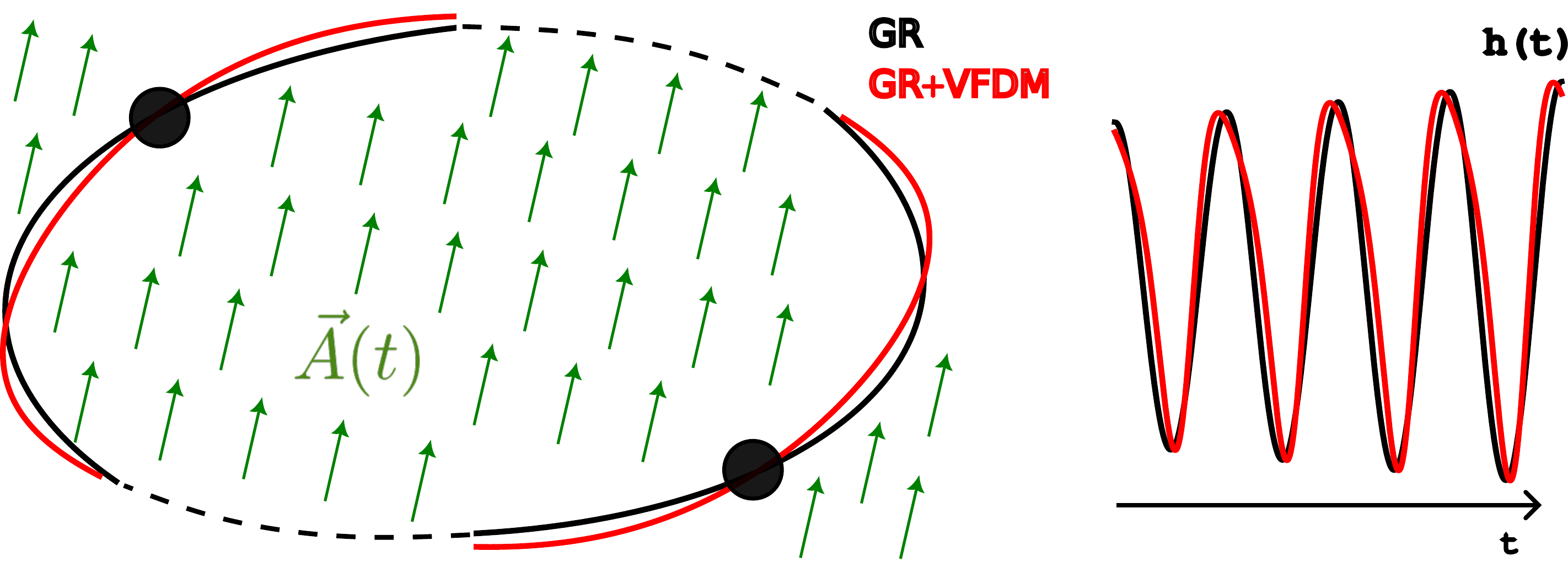}
    \caption{Cartoon illustration of the dark-matter environment effect on the orbit (left) and on the gravitational waves (right) due to an ultra-light vector field (green arrows).}
    \label{fig:intro}
\end{figure}

In this paper, we explore the possibility of detecting such ultralight dark-matter effects (for a vector dark-matter model) through the gravitational waves emitted by massive black hole binaries immersed in such a dark-matter environment and detected (in the future) with the space-borne Laser Interferometric Space Antenna (LISA). Figure~\ref{fig:intro} presents a cartoon illustration of the problem we tackle. We first use post-Newtonian theory to calculate the dark-matter corrections to the orbital binding energy and Kepler's third law. For scalar-field, ultralight, dark-matter models, this effect vanishes for circular orbits, as a consequence of the symmetry of the configuration. For vector-field, ultralight, dark-matter models, however, this symmetry is broken thanks to the anisotropic nature of the vector field. We thus focus on quasi-circular orbits and calculate the vector ultralight dark-matter corrections to the Fourier-domain gravitational-wave phase, using the stationary-phase approximation.  

As we show in the paper, the strength of the ultralight dark-matter effect on the gravitational waveform is proportional to the energy density $\rho_{\rm{DM}}$ of the dark-matter field in the surroundings of the black holes binary. Several works \cite{local_dm_density_1, local_dm_density_2} have estimated the dark-matter density in our Solar System to be $\rho_{\rm{DM}} \sim 0.3 \rm{GeV}/\rm{cm}^3 \sim 0.01 \rm{M}_{\odot}/\rm{pc}^3$. However, the dark-matter density distribution in galaxies is still a matter of debate. In particular, the dark-matter distribution surrounding black holes remains uncertain and varies depending on the dark-matter model. For example, various works \cite{Alcubierre:2024mtq, Davies_2020, Cruz_Osorio_2011, Bar_2019, Aghaie:2023lan, Hancock:2025ois} suggest that scalar and vector field models in the particle regime ($\lambda_c \sim m^{-1} \ll r_s$, with $r_s$ the black hole Schwarzschild radius) \textit{spike} near black holes, where the density may be several orders of magnitude bigger than the one estimated for our Solar System. In our work, we consider fields and black holes that are in the wave regime ($\lambda_c \gg r_s$), where the existence of a local density enhancement is unclear.
We find that the vector ultralight dark-matter corrections are indeed proportional to the dark-matter density, and inversely proportional to a certain power of the ultralight dark-matter mass and the total mass of the binary. Moreover, we find that when the gravitational-wave frequency exceeds a certain threshold, given by a stationary point condition that depends on the vector field mass, the ultralight dark-matter corrections are strongly suppressed. On the other hand, when the gravitational-wave frequency is below this threshold, the vector ultralight dark-matter effect on the gravitational-wave phase for quasi-circular orbits is partially degenerate with the time and phase of coalescence of the binary, rendering this effect hard to measure. This leads to a Heaviside-type modification to the Fourier transform of the gravitational-wave phase, and it is this sudden turn-on/turn-off nature of the modification that yields the dark-matter effect potentially measurable. %We find that, when the vector ultralight dark-matter effect is on, its impact on the gravitational-wave phase for quasi-circular orbits is partially degenerate with the time and phase of coalescence, rendering this effect hard to measure. 

Our work suggests that chirping, massive binaries that sample a range of frequencies in the LISA band (and thus, that cross the stationary point condition activating the Heaviside dark-matter modificatons) are necessary to observe this type of dark-matter effect. Through a Fisher analysis, we estimate that LISA would be able to constraint the ultralight dark-matter mass range $(10^{-19},10^{-16}) {\rm{eV}}$ for gravitational waves emitted by massive [$(10^2,10^6) \rm{M}_{\odot}$] black hole binaries of comparable mass, provided the ultralight dark-matter density is above $(10^{14},10^{15}) \; \rm{M}_\odot \; {\rm{pc}}^{-3}$. %For comparison, the dark-matter density in our Solar System has been estimated to be $\sim 0.3 \rm{GeV}/\rm{cm}^3 = 10^{-2} \rm{M}_{\odot}/\rm{pc}^3$~\cite{local_dm_density_1, local_dm_density_2}, which is many orders of magnitude smaller than what can be probed with LISA. However, the dark-matter density distribution in galaxies is still a matter of debate. In particular, the dark-matter distribution surrounding black holes remains uncertain and varies depending on the dark-matter model. For example, various works \cite{alcubierre2024gravitationalatomstestfield, Davies_2020, Cruz_Osorio_2011, Bar_2019} suggests that scalar field models \textit{spike} near black holes, where the dark-matter density may be many orders of magnitude larger than in our Solar System. 
Our work, therefore, suggests that future LISA observations of the gravitational waves emitted in the quasi-circular inspiral of massive black holes may detect or yield interesting constraints on the density of vector-type ultralight dark matter.

The remainder of this paper presents the derivation of the results summarized above, and it is organized as follows. 
In Sec.~\ref{sec:impact_vector-field dark matter_orbit}, we present the vector-field dark-matter model, while in Sec.~\ref{sec:background-vector-field}, we solve for the evolution of the vector field and its back-reaction onto the perturbations of the spacetime metric. 
In Sec.~\ref{sec:impacto_on_binary}, we show how the presence of a vector-field dark-matter component changes the evolution of the orbital parameters of a binary system.  
In Sec.~\ref{sec:impact_GWform}, we calculate the vector ultralight dark-matter modifications to the gravitational waves emitted by quasi-circular inspiraling binaries, as a consequence of the vector field environment. 
In Sec.~\ref{sec:Fisher_analysis}, we present a Fisher analysis and compute the parameter space of the vector-field dark-matter model that would be detectable given a LISA observation of massive black hole binaries. %Also, in Appendix \ref{sec:multi_scale_analysis}, we present a multiscale analysis of the orbital evolution in a vector-field dark matter environment. %Henceforth, we use the following conventions: we work in geometric units where $m_P = c = 1$, where $m_P$ is the Planck mass; ... {\ny{Insert other conventions you use here (like units)}}
Henceforth, we switch between geometric units (where $G = 1 = c$) and natural units (where $\hbar = 1 = c$), and therefore, we sometimes quote masses in units of length and sometimes in units of energy. The conversion is, of course, simple with, for example, $\tilde{m} = c \, m/\hbar$ in units of length when $m$ is in units of electron-Volts.

%{\ny{Also, I got rid of all the acronyms, because I was getting confused.}}
%%%%%%%%%%%%%%%%%%%%%%%%%%%%%%%%%%%%%%%%%%%%%%%%%%%%%%%%%%%%%%%%%%%%%%%%%%%%%%%%
\section{Ultralight dark-matter vector field around a compact binary} 
\label{sec:impact_vector-field dark matter_orbit}

In this section, we present the vector-field dark-matter model that we study in this paper, following mostly~\cite{LopezNacir:2018epg}, for completeness and to establish notation. 
%In ultralight dark-matter models, the dark matter in virialized halos can be described as a collection of waves with random phases. The occupation numbers in the halos are so high that a classical field approach is applicable. 
Let us then describe the dark matter through an ultralight vector field minimally coupled to gravity, whose action is
\begin{equation}
    \mathcal{S} = \int dx^4 \sqrt{-g} \left[R - \frac{1}{4} F^{\mu\nu} F_{\mu\nu} - \frac{m_A^2}{2} A^{\mu} A_{\mu} + \mathcal{L}_M \right],
    \label{action}
\end{equation} 
where $R$ is the Ricci scalar, $g$ is the determinant of the metric, $\mathcal{L}_M$ is the Lagrangian of the ordinary matter sector, $F_{\mu\nu} = \nabla_{\mu} A_{\nu} - \nabla_{\nu} A_{\mu}$ is the field strength tensor of the massive vector field $A^{\mu}$, with mass $m_A$.  Since we assume that the vector field makes up all the dark matter in the universe, we have a lower bound on the boson mass of $m_A > H_{\rm{eq}} \sim 10^{-28}\rm{eV}$, where $H_{\rm{eq}}$ is the Hubble parameter at the moment of equality between the energy density of the radiation and matter content of the universe.

The equations of motion of the vector field can be derived by varying the action with respect to the field, thus obtaining
\begin{equation}
    \nabla_{\mu} F^{\mu \nu} + m_A^2 A^{\nu} = 0\,,
    \label{eq_motion_vector}
\end{equation}
since there is no direct (non-minimal) coupling of the vector field with the gravitational sector. Similarly, the equations of motion for the metric can be derived by varying the action with respect to the metric, obtaining
\begin{equation}
    {G^{\mu}}_{\nu} = 8 \pi G\, {T_A^{\mu}}_{\nu} +  8 \pi G\, {T_M^{\mu}}_{\nu}\,,
\end{equation}
where $ {G^{\mu}}_{\nu}$ is the Einstein tensor, and ${T_A^{\mu}}_{\nu}$ and $ {T_M^{\mu}}_{\nu}$ are the stress-energy tensor of the vector field and the ordinary matter sector respectively.  We calculate the stress energy of the vector field as usual, $ T^A_{\mu \nu} = -({2}/{\sqrt{g}}) ({\delta \mathcal{S}_A}/{\delta g^{\mu\nu}})$ where $\mathcal{S}_A$ is the action of the vector field, to find
\begin{align}
    {T_A^{\mu}}_{\nu} = F^{\mu \gamma} F_{\nu \gamma} - \frac{{\delta^{\mu}}_{\nu}}{4} F^{\mu\nu} F_{\mu\nu} - m_A^2 (A^{\mu} A_{\nu} + \frac{{\delta^{\mu}}_{\nu}}{2} A^{\mu} A_{\mu} )\,,
    \label{stress_energy_vector_field}
\end{align}
which is symmetric, as expected. 

In this paper, we study the effect of vector dark matter on an inspiraling binary system that emits gravitational-waves in the millihertz band. Therefore, the stress-energy tensor for ordinary matter represents a compact binary, composed of intermediate-mass and supermassive black holes. For simplicity, we assume the binary is in a quasi-circular inspiral orbit and that the black holes are not spinning, leaving extensions of our work to eccentric orbits and spinning black holes for the future.

We consider that the  vector dark matter field around a compact binary can be described as a   classical wave characterized by an amplitude, phase and direction\footnote{We assume that the background vector field is linearly polarized. However, this is not a necessary condition for the local vector field in the halo. For example, in \cite{Amaral:2024tjg,Amin:2022pzv} the authors perform small-scale structure simulations with vector field dark matter with different polarizations for the background field.} that can be considered to be constant within a certain characteristic volume during a characteristic  time interval. We will refer to such space-time volume as a coherent patch. The vector field in different coherent patches can have very different amplitudes, phases and directions. Our model is based on numerical simulations.
In Refs. \cite{Amin:2022pzv,Lopez-Sanchez:2025osk}, the authors study the merger of idealized halos, usually called \textit{solitons}, to characterize small-scale structure formation  with vector field dark matter. As was shown in such spin-1 halo-scale simulations, the system evolves into an approximately spherical configuration, that has a core surrounded by a dark matter density distribution characterized by interference patterns. The characteristic size of the interference granules is given by the de Broglie wavelength of the particle $\lambda_{dB}$. This is similar to what has been shown for scalar field dark matter models  (see for instance Refs. \cite{Schive:2014dra,Veltmaat:2018dfz,Liu:2022rss}). Due to the  evolution of the interference patterns, at a given point in space the field loses coherence after a time  that can be estimated as $t_{coh}\sim \lambda_{dB}/V_0$, where $V_0$ is the local velocity of the dark matter field. These  characteristic scales ($\lambda_{dB}$ and $t_{coh}$) define the typical space-time size of the coherent patches.  

%\tc{citar por papers de uldm solitons: \cite{Schwabe:2016rze, Schive:2014hza}}
We focus on binary systems living inside one of these patches, characterized by a constant dark matter energy density.  Then, a given  binary system is immersed in a vector field environment that is homogeneous over the size of the binary and the gravitational wavelength of the waves emitted by the binary. 
 More concretely, if the binary separation is $r_{12}$ and the orbital velocity is $v_{12}$, then the  wavelength of the gravitational waves is $\lambda_{GW} = r_{12}/v_{12}$, and we require that $r_{12} \ll \lambda_{GW} \ll \lambda_{dB}$, with $\lambda_{dB}$ the local de Broglie wavelenght of the boson. %c\, m/\hbar \equiv \tilde{m}$. %Figure~\ref{} shows a diagrammatic representation of the system we consider in this paper. 
For the ultra-light masses considered in this work $(10^{-19}, 10^{-16}) \; \rm{eV}$ and for the binaries that can be probed by space-borne detectors, the de-Broglie wavelength of the boson  $\lambda_{dB}$ is   much larger than the orbital separation of such compact binary systems, $\lambda_{dB} \gg r_{12}$. Indeed, the de-Broglie wavelength of the boson is%\di{la masa mas chica es $10^{-16} eV$, hay dos ordenes de magnitud,  me parece mas claro tomar esa masa como referencia aca debajo }
\begin{equation}
    \lambda_{dB} \sim 1.3 \times 10^{12}  \left(\frac{10^{-3}}{V_0}\right) \left( \frac{10^{-18}\rm{eV}}{m_A}\right)\rm{km}\,,
    \label{lambda_dB}
\end{equation}
where $V_0$ is the velocity dispersion of the halo, which can be estimated from the Virial velocity, e.g.~$V_0 \sim 10^{-3}$ for the Milky Way. On the other hand, for the systems that can be probed by LISA with a high signal-to-noise ratio, such as supermassive black holes of $M\sim10^6 \rm{M}_{\odot}$, the largest separations correspond\footnote{In principle, LISA can detect binaries with lower masses at much larger separations (which would merge in the  band of ground-based detectors), but at much lower signal-to-noise ratios, since the latter scales with the total mass.} to the lowest frequencies LISA is sensitive to, $f \sim 10^{-5}\rm{Hz}$. Using Kepler's third law we can calculate the semi-major axis of such systems, giving  
\begin{equation}
    r_{12} \sim 3.2\times10^{8}\left(\frac{M}{10^6 \rm M_{\odot}}\right)^{\frac{1}{3}}\left(\frac{f}{10^{-5}\rm Hz}\right)^{-\frac{2}{3}} \rm{km}\,,
    \label{typical_radius}
\end{equation}
and for the corresponding emitted gravitational wave $\lambda_{GW} \sim 10^{10}\,\rm{km}$. Furthermore, for $m_A < 10^{-16}\rm{eV}$, we have that 
\begin{equation}
    t_{coh} \sim 65 \,\rm{yrs} \left( \frac{10^{-3}}{V_0}\right)^2 \left(\frac{10^{-18}\rm{eV}}{m_A} \right) \,, 
\end{equation}
so for the field masses considered in this work the patch remains coherent during a typical LISA observation, which ranges from 1 yrs to 4 yrs. In summary, we have that the de-Broglie wavelength is much larger than the orbital separation of the binary, and during the observation, the patch remains coherent. We therefore expect the background vector field (and its associated metric perturbation) to be approximately \textit{homogeneous} at scales comparable to the size of the binary system, i.e.~$\bar{A}^{\mu}$ and $h^{A}_{\mu \nu}$ are approximately independent of spatial coordinates and only depend on time. In Appendix \ref{appendix:backreaction} we give a more detailed discussion of the previous assumption, based on numerical simulations.

We assume that the binary system and the vector field produce small departures from a background vacuum metric, both of which can be treated perturbatively. Thus, we start by splitting the metric and the vector field as
\begin{align}
    g_{\mu\nu} &= \bar{g}_{\mu\nu} + h^{A}_{\mu\nu} + h^{M}_{\mu\nu}\,, \\
    A_{\mu} &= \bar{A}_{\mu} + \delta A_{\mu}\,,
\end{align}
where $\bar{g}_{\mu\nu}$ and $\bar{A}_{\mu}$ are the background metric and vector field respectively, while the metric perturbations $h^{A}_{\mu\nu}$ and $h^{M}_{\mu\nu}$ are produced by the vector field and the binary system respectively. The quantity $\delta A_{\mu}$ is a small perturbation that we will neglect henceforth, but that in principle includes inhomogeneities and the backreaction of the binary on the vector field. We here ignore the cosmological expansion of the universe and assume the background metric is the flat Minkowski spacetime\footnote{The typical time scales of a gravitational-wave detector ($\sim$ yrs) are much smaller than the Hubble time in the matter era, and the main correction due to cosmic expansion is a redshift in the wavelength of the propagating gravitational wave \cite{Bieri:2017vni}.}, $\bar{g}_{\mu\nu}  = \eta_{\mu\nu}$. Thus, when we ignore the presence of the vector field, the metric perturbation $h^{M}_{\mu\nu}$ is nothing but the post-Newtonian solution for an inspiraling binary, as reviewed for example in \cite{Blanchet:2013haa}. 

With this set-up, we wish to study the effect of the vector field metric perturbation $h^{A}_{\mu\nu}$ on the  metric perturbation induced by the binary system $h^{M}_{\mu\nu}$, and we do so by expanding  the field equations (Einstein plus vector field) in both perturbations.
Given the above, the set-up of the problem lends itself perfectly to the technique of separation of scales when solving the perturbed field equations. 
We now proceed to solve the field equations with the assumptions described above. First, we work at the largest scales and solve for (i) the vector field, and (ii) the metric perturbation induced by the vector, in both cases ignoring the ordinary matter sector. That is, we solve first
\begin{equation}
    \label{eq:A-EoM-bg}
    \nabla_{\eta} F^{\mu \nu}[\bar{A}_\alpha] + m_A^2 \bar{A}^{\nu} = 0\,,  
\end{equation}
where $\nabla_{\eta}$ is the covariant derivative with respect to the Minkowski metric and $F^{\mu \nu}[\bar{A}_\alpha]$ is the field strength tensor associated with the background vector field. This is an equation of ${\cal{O}}(\bar{A}^{\mu})$, which gives us a background solution for the vector field $\bar{A}^{\mu}$. Given this solution, we then solve 
\begin{equation}
    \label{eq:G-A}
    G_{\mu \nu}[h_{\alpha \beta}^A] = 8 \pi \,T_{\mu \nu}^A[\bar{A}_{\alpha}]\,,
\end{equation}
where $G_{\mu \nu}[h_{\alpha \beta}^A]$ is the part of the linearized Einstein tensor that depends on $h_{\alpha \beta}^A$, while $T_{\mu \nu}^A[\bar{A}_{\alpha}]$ is the part of the linearized stress-energy tensor that depends on $\bar{A}^{\mu}$. This is an equation of ${\cal{O}}(\bar{A}^{\mu} \bar{A}_{\mu})$, which gives us a solution for the metric perturbation $h_{\mu \nu}^A$ induced by the vector field $\bar{A}_{\alpha}$. 

With this in hand, we then focus on the metric perturbation induced by the ordinary matter sector. In the absence of the vector field, the solution for $h_{\mu \nu}^M$ can be obtained through standard post-Newtonian methods~\cite{Blanchet:2013haa}, as discussed earlier. We then need to understand the backreaction of the vector field $\bar{A}^{\mu}$ on $h_{\mu \nu}^M$. The dominant correction to this metric perturbation is that induced by the correction to the trajectories of the binary system, which, as we will see, is effectively a conservative force induced by $\bar{A}^{\mu}$ that modifies the binary's binding energy. By the balance law, this modification then alters the evolution of the orbital frequency, and thus, the gravitational-wave phase. This is the main and largest effect we consider in this paper, and thus, we leave the full back-reaction of $h_{\mu \nu}^A$ on $h_{\mu \nu}^M$ and vice versa to future work (see Appendix \ref{appendix:backreaction} for a discussion).   

%%%%%%%%%%%%%%%%%%%%%%%%%%%%%%%%%%%%%%%%%%%%%%%%%%%%%%%%%%%%%%%%%%%%%%%%%%%%%%%%
\section{Evolution of the background ultralight dark-matter vector field and its metric perturbation} 
\label{sec:background-vector-field}

Following the considerations in the previous section, we model the vector dark matter field in one coherent patch as a homogeneous vector field $\bar{A}^{\mu}$, and calculate its associated metric perturbation. The dynamics of the background vector field is given by Eq.~(\ref{eq:A-EoM-bg}). From this equation, we can extract a constraint equation for the temporal component and three equations of motion for the spatial components. As we discussed earlier, we can assume that the background vector field is spatially homogeneous, $\bar{A}^{\mu} = \bar{A}^{\mu}(t)$. Under this assumption, the temporal component of the background vector field exactly vanishes  $\bar{A}^0 = \bar{A}_0 = 0$, while the spatial components satisfy the equations of motion
\begin{equation}
    \ddot{\bar{A}}_i + m_A^2 \bar{A}_i = 0\,,
\end{equation}
where the overhead dots stand for time derivatives. This equation of motion can be solved as a harmonic oscillator for each component,
\begin{equation}
    \bar{A}^i(t) \sim \tilde{A} \, \cos(m_A t+ \gamma)\,\hat{A}^i\,,
\end{equation}
where $\tilde{A}$ is a constant amplitude, $\gamma$ is a constant phase, both determined by the initial conditions, and $\hat{A}^i$ is a unit vector pointing in the direction of the vector field. %{\ny{Dianna tiene razon.}} 

With this background vector field, we can now search for solutions to the metric perturbation it generates, which is given by Eq.~\eqref{eq:G-A}. %Since the background vector field is homogeneous, so are its associated metric perturbations. 
In the binary's source frame, the metric perturbations can be decomposed in Newtonian gauge as
\begin{equation}
    h^{A}_{00} = 2\phi\,,  \quad h^{A}_{0i} = B_i\,, \quad h^{A}_{ij} = - 2 \psi \delta_{ij} + 2 E_{ij}\,,
\end{equation}
%{\ny{Ojo, que dijimos que todo dependia del tiempo solamente, asi que no pueden aparecer gradientes espaciales en la equacion de abajo (en la de arriba esta ok, si queres, pero chequea.}}
where $(\phi,\psi, B_i, E_{ij})$ are metric perturbation functions of time only, which satisfy ${B^{i,}}_i = 0$ and ${E^i}_{j, i} = {E^i}_i = 0 $. 
At linear order in the metric perturbations, the spatial components of Eq.~\eqref{eq:G-A} then becomes
\begin{align}
    2 \ddot{\psi} \,{\delta^i}_j+{\ddot{E}^i}_{\,\,j}- \partial^{(i} \dot{B}_{j)}  = 8\pi G \,&{{T_A}^{i}}_{j}[\bar{A}^{\mu}] \,, \label{einstein_SS}
    %2 \ddot{\psi} \,{\delta^i}_j+{\ddot{E}^i}_{\,\,j} = 8\pi G \,&{{T_A}^{i}}_{j}[\bar{A}^{\mu}] \,, \label{einstein_SS}
\end{align}
% \begin{subequations}
% \begin{align}
%     \partial_i \partial^i \psi = 4\pi G \,&{{T_A}^{0}}_{0}[\bar{A}^{\mu}]\,, \\[5pt]
%      ({\delta^i}_j\partial_i\partial^i - \partial^i\partial_j)(\phi-\psi)\qquad& \nonumber\\ 
%     +2 \ddot{\psi} \,{\delta^i}_j+{\ddot{E}^i}_{\,\,j} - \partial_i\partial^i {E^i}_j 
%     - \dot{B}_{(i,j)}  = 8\pi G \,&{{T_A}^{i}}_{j}[\bar{A}^{\mu}] \,, \label{einstein_SS}\\[5pt]
%     \partial_i\partial^i B_j - 4 \,\partial_j \dot{\psi} = 16 \pi G \,&{{T_A}^{0}}_{i}[\bar{A}^{\mu}]\,,
% \end{align}
% \end{subequations}
where the energy momentum tensor on the right-hand side is the part of the linearized stress-energy tensor that depends on the background vector field.
% {\ny{You still have a spatial gradient above, and you said the fields only depended on time...}} \tc{Done}

By considering a homogeneous field in flat spacetime we can further simplify the stress-energy tensor given in Eq.~(\ref{stress_energy_vector_field}). First, we note that $-\frac{1}{2} F^{\mu\nu} F_{\mu\nu} + m_A^2 A^{\mu} A_{\mu} = -\dot{A}^2 + m_A^2 A^2$ and $F^{\mu \gamma} F_{\nu \gamma} = \dot{A}^2$, where the square means the spatial norm in flat spacetime. Then, the components of the energy-momentum tensor for the homogeneous field are
\begin{subequations}
    \begin{align}
        {{T_A}^{0}}_{0} &= - \frac{m_A^2 \tilde{A}^2}{2} \equiv -\rho_A\,, \\[5pt]
        {{T_A}^{0}}_{j} &= 0\,, \\[5pt]
        {{T_A}^{i}}_{j} &= \rho_A \cos(2 m_A t+\gamma) \, {\hat{X}^i}_j \equiv  P_A \,{\delta^i}_j + {\Sigma_A^i}_{j} \,,
    \end{align}
\end{subequations}
where ${\hat{X}^i}_{\,\,j} = {\delta^i}_j - 2\hat{A}^i\hat{A}_j$. We have identified the time-time component of the stress-energy tensor as the energy density $\rho_A$ of the vector field in the local patch, which on the timescales considered here, is approximately constant. The space-space components can be split into a trace part, given by the pressure of the field $P_A$, and a traceless part ${\Sigma_A^i}_j$, which can be identified as the shear of the field. The presence of the shear in the energy-momentum tensor is a main difference with respect to ultralight scalar field models. In particular, the shear is responsible for inducing a non-vanishing effect on circular orbits, in contrast with scalar models, where the effect vanishes due to the symmetry of the configuration.

Now we return to the Einstein equations. As will become clear in the next section, to study the impact of the vector-field dark matter on the trajectories of the binary system, and thus, the gravitational-wave metric perturbation, we need Eq.~(\ref{einstein_SS}). With the above stress-energy tensor, this equation simplifies to 
\begin{equation}
    -2  \ddot{\psi}\,{\delta^i}_j - \partial^{(i} \dot{B}_{j)} 
    - \frac{1}{2} {\ddot{E}^i}_{\,\,j} = 8\pi G \,\rho_{A}  \cos(2 m_A t + \gamma) {\hat{X}^i}_{\,\,j}\,.
    \label{einstein_SS_homogeneous}
\end{equation}
As we will show in Sec.~\ref{sec:impact_GWform}, the energy density of the field controls the magnitude of the correction to the trajectories of the binary system, and thus, the gravitational-wave phase evolution. 

%\noindent where we have assumed homogeneous fields ($k\to0$). %and where we focus on time scales such that the expansion of the universe is negligible ($H \sim 0$, with $H$ the Hubble parameter). %Also, for an homogeneous DM field we have that $\partial^{i} S_{j} = 0$. 

%%%%%%%%%%%%%%%%%%%%%%%%%%%%%%%%%%%%%%%%%%%%%%%%%%%%%%%%%%%%%%%%%%%%%%%%%%%%%%%%
\section{Impact of Ultralight Dark Matter Vector field on the Orbital Motion of a Compact Binary} \label{sec:impacto_on_binary}

\begin{figure}[th]
    \centering
    \includegraphics[width=0.8\linewidth]{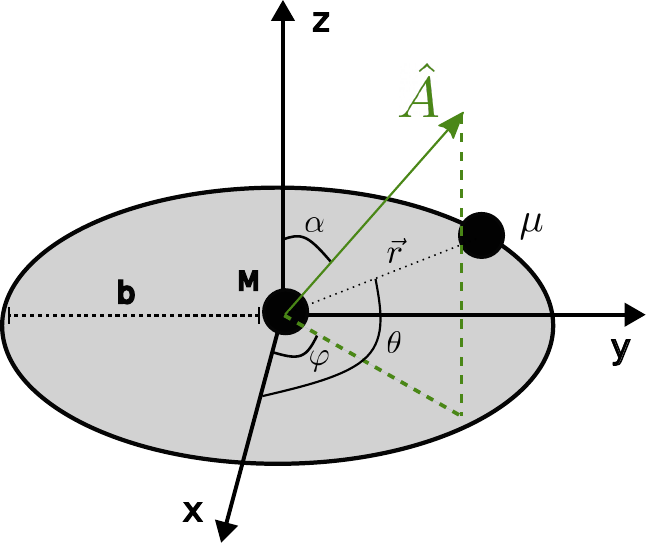}
    \caption{Diagram of the effective one-body problem. The origin of coordinates is placed at the center of mass of the binary, and the position of the reduced mass is described by the vector $\vec{r} = b\left[\cos(\theta)\, \hat{x} + \sin(\theta)\, \hat{y}\right]$. The green arrow points in the direction of the vector field, which is defined with the angles $\varphi$ and $\alpha$.}
    \label{fig:fig_kepler}
\end{figure}

In this section we consider the impact of the oscillating gravitational potentials governed by Eq.~(\ref{einstein_SS_homogeneous}) on the binary orbit. We start by considering two masses bound gravitationally in a circular orbit of radius $R(t)$, neglecting gravitational wave back-reaction. We place the origin of coordinates at the center of mass of the system, and transform the problem into an effective one-body one, where a point-particle (or test mass) with mass equal to the reduced mass $\mu$ of the binary is in orbit around a gravitating object with mass equal to the total mass of the binary $M$. The position of the reduced mass particle is denoted by $r^i(t) = r^i_1-r^i_2$, where $r^i_1$ and $r^i_2$ are the positions of the two masses in the real problem. The binary system is immersed in a vector field dark-matter background, which is approximately homogeneous at scales of $\mathcal{O}(|r^i|)$, but time dependent. The field is pointing in an arbitrary direction (parameterized in terms of two angles, $\alpha$ and $\varphi$) with respect to the orbital plane, which we assume to be constant in time. For an illustration of the configuration see Fig.~\ref{fig:fig_kepler}. 

%\begin{figure}[H]
%    \centering
%    \includegraphics[width=0.8\linewidth]{Figures/figura_sistema.pdf}
%    \caption{Illustration of the two body problem considered in this work.}
%    \label{fig:figura_sistema}
%\end{figure}

The oscillating potentials induced by the vector-field dark-matter environment causes small deviations in the system's orbit. These deviations can be captured in general relativity through the vector-field corrections to the \textit{geodesic deviation equation}. In other words, in the effective-one-body problem, the test particle follows a geodesic (neglecting gravitational-wave backreaction) in the geometry of the larger body, but this geodesic is perturbed by vector-field effects; the perturbations to the geodesic motion can be captured through the geodesic deviation equation. In the weak-field approximation and using Fermi normal coordinates, the geodesic deviation equation for the reduced mass reads 
\begin{equation}
    \frac{d^2 r^{i}}{d t^2} = \left({R^{i}}_{0j0}[h^A_{\mu\nu}]+{R^{i}}_{0j0}[h^M_{\mu\nu}]\right) r^{j}\,.
    \label{geodesic_eq}
\end{equation}
to first order in metric perturbations, where ${R^{i}}_{0j0}[h^A_{\mu\nu}]$ and ${R^{i}}_{0j0}[h^M_{\mu\nu}]$ are the components of the linearized Riemann tensor associated with the metric perturbations generated by the vector field and the binary system respectively. 

The contribution from the binary system to the linearized Riemann tensor in Eq.~(\ref{geodesic_eq}), ${R^{i}}_{0j0}[h^M_{\mu\nu}]$, yields the post-Newtonian equations of motion in the center of mass frame. In particular, when the metric perturbation $h_{\mu \nu}^M$ is taken to leading post-Newtonian order, one obtains the usual Newtonian law of gravity, i.e.~${R^{i}}_{0j0}[h^M_{\mu\nu}] r^j = -({G\,M}/{r^2}) r^i$ (see for instance Sec.~17.4 in~\cite{Misner:1973prb}). The contribution from the vector field can then be interpreted as a perturbation to the post-Newtonian equations of motion. The components of ${R^{i}}_{0j0}[h^A_{\mu\nu}]$ can be written in terms of the vector field fluid variables using Eq.~(\ref{einstein_SS_homogeneous}), as
\begin{align}
    {R^{i}}_{0j0}[h^A_{\mu\nu}] &= -{\delta^i}_j \, \ddot{\psi}  - \frac{1}{2} {\ddot{E}^i}_{\,\,j} + \frac{1}{2} \partial^{(i} \dot{B}_{j)} \\[5pt]
    &= -4\pi \rho_{A}  \cos(2 m_A t + \gamma) {\hat{X}^i}_{\,\,j}\,.
\end{align}
 %The different components of the force are given in Eqs.~\ref{vector_force_components}.
Inserting the previous expression in Eq.~(\ref{geodesic_eq}), we obtain the modified Newtonian equation for the binary system in the vector-field environment,
\begin{equation}
    \vec{a} + \frac{G M}{r^2} \, \hat{r} = - 4\pi\rho_{A} r\,\cos(2 m_A t + \gamma) [\hat{r} - 2 (\hat{r}\cdot \hat{A}) \hat{A}]\,,
    \label{eq_newton_binary}
\end{equation}
where $\vec{a}$ is the relative acceleration between the bodies and we have here kept the post-Newtonian equations of motion to leading order only (and moved the Newtonian gravitational acceleration to the left-hand side)

The effect of the vector-field environment on the orbit can be interpreted as a perturbation to the usual Newtonian law of gravity. We thus identify the right-hand side of Eq.~\eqref{eq_newton_binary} as a perturbative force per mass unit, that the vector field exerts over the binary system. The corresponding produced acceleration can then be written as
\begin{equation}
    f^i = - 4\pi\rho_{A}  \cos(2 m_A t + \gamma) \, r^j {\hat{X}^i}_{\,\,j}\,,
    \label{vector_force}
\end{equation}
where we recall that ${\hat{X}^i}_{\,\,j} = {\delta^i}_j - 2\hat{A}^i\hat{A}_j$. Moreover, since the unperturbed problem reduces to Newtonian gravity, we know that Kepler's laws are valid to leading order, so the frequency of the orbit is $\omega^2 \sim M/r^3 + {\cal{O}}(\rho_A)$, and the binding energy of the orbit is the usual Newtonian potential, namely $E_M \sim -G\mu M/2r + {\cal{O}}(\rho_A)$.

The acceleration in Eq.~\eqref{vector_force} can be derived from the following time-varying potential,
\begin{equation}
    V_A = 2\pi \mu \rho_A \cos(2 m_A t + \gamma) \left[ \vec{r}^{\,\,2} - 4 \,(\vec{r}\cdot \hat{A})^2\right]\,,
    \label{binding_energy_A}
\end{equation}
where $f_i = -\partial_i V_A$. 
%{\ny{Please check whether there is a + or a - sign here.}} \tc{it is a -}.
This potential is invariant under time reversal\footnote{The phase angle $\gamma$ is set by the initial conditions to the solution of the vector field equation of motion. In particular, $\gamma$ depends on the initial velocity of the vector field, which flips sign under time-reversal. Therefore, to keep the initial conditions invariant under time reversal, then $\gamma \to -\gamma$ when $t \to -t$. With this, under time reversal $\cos(2 m_A t + \gamma) \to \cos(-2 m_A t - \gamma) = \cos(2 m_A t + \gamma)$.}, and thus, the force associated with it is conservative. 
% We thus have that the vector field exerts a conservative force over the binary system. 
The binding energy of the system is then modified to be $E = E_M + V_A$. The perturbation $V_A$, despite being time varying, does not change the binding energy, on average over time scales of order $m_A^{-1}$ or $1/\omega$, with $\omega$ the angular frequency of the orbit. In this paper, we consider binary systems in the LISA band, over an observation period of order a yr, so the binding energy is not modified on average. However, as we show in Sec.~\ref{sec:impact_GWform}, this perturbation does impact the gravitational waveform.
% whose frequency resonates with the vector field at some point in the chirp of the frequency. We thus have that, in the period of observation considered here ($\sim$ yrs for LISA), the binding energy is not modified on average. However, as we show in Sec.~\ref{sec:impact_GWform}, this perturbation has an impact on the gravitational waveform.

%{\ny{Your formatting of equations is very strange, because you start a new paragraph and then you don't indent. you don't have to do that. i've corrected it throughout. }}

% %------------------------------------------------------
% \subsection{Perturbed Kepler problem}

The perturbation in the binding energy $V_A$ induces a perturbation in the evolution of the semi-major axis of the orbit, given by the oscillating potentials generated by the vector field. Let us now calculate this variation with the \textit{method of osculating orbits}, which is ideally suited to perturbations of Keplerian orbits. This method gives the equations of motion for the parameters that are otherwise conserved in the usual Kepler problem. 
To calculate the variation in the semi-major axis $b$, it is convenient to decompose the  acceleration  in a polar basis, given by $\{\hat{r}, \hat{\theta}, \hat{z}\}$, as 
\begin{equation}
    \Vec{f} = f_r\, \hat{r} + f_{\theta}\, \hat{\theta} + f_z \hat{z}\,,
\end{equation}
where $\hat{z}$ is normal to the orbital plane.  In this basis, one can show that the equations of motion for the semi-major axis of a circular orbit is given by\footnote{For a detailed derivation of the osculating equations, see for instance Sec.~3.3 of Ref.~\cite{Poisson_Will_2014}.}
\begin{equation}
    \dot{b} = \frac{2}{\omega} f_{\theta}\,, \label{adotovera}
\end{equation}
where $\omega = 2 \pi/P = \sqrt{G M/b^3}$ by Kepler's third law, and $P$ is the orbital period. 

The method of osculating orbits also predicts variations in other orbital parameters due to the presence of the vector field environment, such as a variation in the orbital eccentricity, or in the relative inclination angle $\iota$ between the orbit plane and the observation plane. However, we here neglect such effects and leave them for future work for the following reasons. The variation in the angle $\iota$ can be neglected because it only impacts the gravitational wave amplitude (for non-spinning binaries), which in general is harder to measure in comparison to perturbations in the gravitational-wave phase (such as the contribution from $\dot{b}$ we consider here). Moreover, general relativity is very efficient at circularizing orbits through gravitational-wave emission, so in the regime where the vector field effect is much smaller than those in general relativity, it is a good approximation to work in the quasi-circular limit, and we can neglect any dark-matter excitation of the eccentricity. 

The equation of motion for the semi-major axis (Eq. \ref{adotovera}) deserves further discussion. First, we see that, for circular orbits, we have a non-zero perturbation of the orbit, as long as the perturbing forces have tangential components. Scalar field models do not have tangential components due to the symmetry of the field. Fields of higher spin, however, can break this symmetry, thus producing a non-vanishing effect, even for circular orbits. Note also that the perturbing force is conservative, so the temporal evolution of the semi-major axis is also conservative and, in particular, it is not driven by dissipative forces. Gravitational-wave back-reaction is a dissipative effect that also induces a time-variability in the semi-major axis, forcing the binary to inspiral, but this back-reaction effect is qualitatively different from the conservative effect of the perturbing vectorial force studied here.   

%For simplicity we assume circular orbits, so $r^j = b\, \hat{r}^j$ and $e = 0$. 
Let us now calculate the components of the force using Eq.~(\ref{vector_force}). To do this, it is convenient to write the vector field in spherical polar coordinates, as $\hat{A} =  \sin(\alpha)\cos(\varphi) \, \hat{x} + \sin(\alpha)\sin(\varphi) \,\hat{y} + \cos(\alpha) \, \hat{z}$, where $\hat{x}$-$\hat{y}$ is the orbital plane. The position of the reduced mass on the orbital plane in polar coordinates can be then written as $\vec{r} = b\left[\cos(\theta)\, \hat{x} + \sin(\theta)\, \hat{y}\right] + z \hat{z}$. For a diagram of the configuration, see Fig.~\ref{fig:fig_kepler}. With this in hand, we obtain the tangential component of Eq.~(\ref{vector_force}) at $z=0$,
%\begin{subequations}
\begin{align}
    %\left(f_{r}\right)_A &= -4\pi\rho_{A} b \cos(2 m t + \gamma) \left(\cos(2 \theta - 2 \varphi) \sin(\alpha)^2 - \cos(\alpha)^2\right)\,,\\
    \left(f_{\theta}\right)_A &=  8 \pi\rho_{A}\, b\, \cos(2 m_A t + \gamma)\sin(2 \varphi - 2 \theta) \sin(\alpha)^2\,. \label{force_theta} %\\
    %\left(f_{z}\right)_A &=  4 \pi\rho_{A}\, b\, \cos(2 m_A t + \gamma)\sin(\theta - \varphi) \sin(2\alpha)^2\,.
\end{align}
%\label{vector_force_components}
%\end{subequations}
%where $\hat{A}_i = A_i/A$.
Observe that the force is proportional to the dark matter density $\rho_A$, as expected. Observe also that the force vanishes if the vector is exactly orthogonal to the orbital plane, since then the vector field would not exert any tangential force. 
%{\ny{please check that you agree.}} \tc{I agree.}

% {\ny{I have a question here. Did you check that there are no other components of the vectorial force in the circular limit? That is, I could imagine that the vector A may induce a force that induces precession of the test-particle off of the initial orbital plane, i.e. a change in iota. Does this not happen?}} {\ny{You still need to address this comment.}} \tc{I am confused, are't we talking about the other effects two paragraphs above, in the previous comment?}

We can now calculate the equation of motion of the semi-major axis by inserting the vector-field force [Eq.~\eqref{force_theta}] in the Keplerian expression for $\dot{b}$, given in Eq.~(\ref{adotovera}). By considering the conservative contribution from the vector-field environment, we obtain that the variation in the semi-major axis is
\begin{align}
    %\frac{\dot{b}}{b} = &-\frac{16}{15}\frac{G^3}{c^5}\frac{\eta M^3}{b^4}
    %+\frac{16 \pi G \rho_A}{w}\,\cos(2mt+2\gamma) \times\label{adotoveravec}\\[3pt]
    %&\left[ (\hat{A}_x^2 - \hat{A}_y^2) \sin(2 w (t-t_0)) + 2 \hat{A}_x \hat{A}_y  \cos(2 w(t-t_0)) \right]\,.\nonumber
    \left(\frac{\dot{b}}{b}\right)_A = \frac{16 \pi G\rho_A}{c^2 \omega}  \cos(2 m_A t + \gamma) \sin( 2 \varphi- 2 \omega t) \sin(\alpha)^2\,. \label{adotoveravec}
\end{align}
%where $\eta = \mu/M$ is the symmetric mass ratio. 
As mentioned earlier, the perturbation to the semi-major axis vanishes when the vector field is normal to the orbit plane, due to the symmetry of the circular orbit. Observe also that the perturbation is greatest when the vector is parallel to the orbital plane. 
% In the next section, we use the previous expression to calculate the dephasing in the gravitational wave due to the vector field environment.

%{\ny{Please check the mental math below and put in the factors of G and c.}} \tc{Done.}

How large is this perturbation relative to other perturbations we expect the binary system to experience? The main other perturbation is gravitational-wave backreaction, which induces a change in the semi-major axis of the form~\cite{Peters:1964zz}
\begin{align}
    \left(\frac{\dot{b}}{b}\right)_{M} = - \frac{64}{5} \frac{G^3\eta}{c^5 b} \left(\frac{M}{b}\right)^3\,,
    \label{eq:peters}
\end{align}
where $\eta \equiv \mu/M$ is the symmetric mass ratio.
The  minus sign here indicates the well-known result that gravitational waves induce an inspiral decay of the orbit until merger. The vector-field perturbation considered here induces a similar inspiral effect, therefore enhancing the rate of inspiral and accelerating the merger (although $\dot{b}_A$ oscillates, and thus its sign can change with time). To leading order in vector field and gravitational-wave perturbations, the total decay rate of the semi-major axis is simply the sum of Eqs.~\eqref{adotoveravec} and~\eqref{eq:peters}. 

A comparison of the rate of change of the semi-major axis induced by the vector field and by gravitational waves is instructive. We can take the ratio of Eqs.~\eqref{adotoveravec} and~\eqref{eq:peters} to find
\begin{align}
      &\frac{\left({\dot{b}}/{b}\right)_{A}}{\left({\dot{b}}/{b}\right)_{M}} \sim \frac{c^3 b^4}{G^2 \eta M^3 \omega} \rho_A \label{eq:ratio_bdots}\\
      &\quad\sim 1.5\times 10^{-16}\frac{\eta^{-1}\rho_A }{0.3\,\frac{\mathrm{GeV}}{\mathrm{cm^3}}}\left(\frac{10^{6}\mathrm{\rm{M}_{\odot}}}{M}\right)^{5/3}\left(\frac{10^{-5}\mathrm{Hz}}{f}\right)^{8/3}\,,\nonumber
\end{align}
where we used the solar system dark matter energy density as the reference value for $\rho_A$.
From this ratio, we observe that the effects of the vector field are enhanced for asymmetric binaries that are widely separated, as we anticipated in the introduction.

%%%%%%%%%%%%%%%%%%%%%%%%%%%%%%%%%%%%%%%%%%%%%%%%%%%%%%%%%%%%%%%%%%%%%%%%%%%%%%%%
\section{Impact on Gravitational Waves} \label{sec:impact_GWform}

In this section we calculate the impact of the vector-field dark-matter environment on the gravitational waves emitted by a quasi-circular inspiralling compact binary. In Sec.~\ref{sec:spa_GR}, we review the calculation of the gravitational waveform in general relativity using the stationary-phase approximation. We also show how to perturb the gravitational waveform using the previous formalism. In Sec.~\ref{sec:calculation_dephasing}, we calculate the dephasing of the gravitational wave due to the vector-field dark-matter environment.

%------------------------------------------------------
\subsection{The stationary-phase approximation}\label{sec:spa_GR}

%In this section we review the \textit{stationary phase approximation} (SPA) in general relativity for the calculation of the gravitational-wave phase. 
We start by considering a linear perturbation of the metric around a Minkowski background as $g_{\mu\nu} = \eta_{\mu\nu} + h^M_{\mu\nu}$, with $|h^M_{\mu\nu}|\ll1$. In general relativity, it is well known that gravitational waves in the far field can be expressed in a multipolar expansion. For a binary system and to leading-order in post-Newtonian theory, one can write  
% , for a binary system in the non-relativistic limit, the amplitude of the gravitational wave emitted can be calculated with a multipolar expansion of the stress-energy tensor of the binary. At leading order in such expansion, the amplitude can be written in terms of the mass-quadrupole moment of the system as
\begin{equation}
    h_{ij}^{M, TT} = \frac{2 G}{c^4 D}\,\ddot{M}_{ij}^{TT}\,,%\Lambda_{ij,kl}(\hat{n}) \ddot{M}^{kl}\,,
\end{equation}
where $D$ is the distance from the source to the detector in the source frame, the super-script $TT$ indicates the transverse-traceless part of the tensor, and $M_{ij}$ is the mass-quadrupole moment of the binary, with overhead dots standing for time derivatives. For a circular orbit of two point masses, the mass-quadrupole moment is given by
\begin{align}
    \ddot{M}_{11} &= 2 \mu (\pi f b)^2 \cos(2\pi f t)\,,\\[3pt]
    \ddot{M}_{22} &= 2 \mu (\pi f b)^2 \sin(2\pi f t)\,,
\end{align}
where $f$ is the gravitational-wave frequency, which is related to the orbital angular frequency via $\omega = \pi f$. 

Gravitational waves in general relativity have only two polarizations, denoted $h_+$ and $h_\times$. One can thus write
\begin{equation}
    h_{ij}^{M, TT} = h_+ \begin{pmatrix}
1 & 0 & 0 \\
0 & -1 & 0 \\
0 & 0 & 0
\end{pmatrix} + h_{\times} \begin{pmatrix}
0 & 1 & 0 \\
1 & 0 & 0 \\
0 & 0 & 0
\end{pmatrix}\,.
\end{equation}
%\noindent where $\Lambda_{ij,kl} \equiv P_{il} P_{jm} - \frac{1}{2} P_{ij} P_{lm}$, with $P_{ij} = (\gamma_{ij} - \hat{n}_i\hat{n}_j)$, is a transverse-traceless projector.
In this paper, we focus on the combination $h \equiv h_+ - i h_{\times}$. This combination can be parameterized as a product of a slowly varying amplitude $\mathcal{A}(t)$ and a rapidly-oscillating phase $2\Phi(t)$ via 
\begin{equation}
    h(t) = \mathcal{A}(t)\,e^{2i\Phi(t)}\,.
\end{equation}
Moreover, in this paper we will particularly focus on the Fourier transform of $h$, because most data analysis and parameter estimation for ground-based gravitational-wave detectors is carried out in the frequency domain. In particular, we here use the convention
\begin{equation}
    \tilde{h}(f) = \int \mathcal{A}(t) e^{i(2\Phi(t) - 2\pi f t)}\,dt\,,
    \label{h_of_f_ini}
\end{equation}
where the overhead tilde stands for the Fourier transform. 

We can evaluate the Fourier transform with the stationary-phase approximation. In this approximation, we Taylor-expand the exponent in the integrand of the Fourier transform around $t=t_{*}$, where $t_{*}$ is a ``stationary point'' where $\dot{\Phi}(t_{*}) = \pi f$.
%, where the subscript $*$ indicates that the phase is evaluated at $t=t_{*}$. 
Then, we have that
\begin{align}
    (2\Phi(t) - 2\pi f t) &\sim 2\Phi(t_*) - 2\pi f t_{*} + \ddot{\Phi}\big|_{t_*} (t-t_{*})^2\,, 
    \\
    \mathcal{A}(t) &\sim \mathcal{A}(t_{*})\,. %+ \dot{A}\big|_{t_{*}} (t-t_{*})
\end{align}
We can now solve the integral in Eq.~(\ref{h_of_f_ini}) by integrating the remaining Gaussian to obtain
\begin{equation}
    \tilde{h}(f) = i \tilde{\mathcal{A}}(t_{*}) \sqrt{\frac{\pi}{\ddot{\Phi}_*}} e^{i(2\Phi_* - 2\pi f t_*)} \equiv \tilde{\mathcal{A}}(f)\,e^{i\Psi(f)}\,.
\end{equation}

Since $\tilde{\mathcal{A}}(f)$ is a slowly varying amplitude, $\tilde{h}(f)$ is controlled by the phase $\Psi(f)$. For gravitational-wave interferometers, in fact, the Fourier phase is one of the most important elements of the waveform model. We can write the Fourier phase using that
\begin{align}
    \Phi(t_*) &= \Phi_c -  2\pi \int^{\infty}_{f/2} \frac{F(F')}{\dot{F}(F')}dF'\,, \\[5pt]
    t_* &= t_c - \int^{\infty}_{f/2} \frac{1}{\dot{F}(F')} dF',
\end{align}
where $F'$ is the integration variable, while $F = \omega/(2 \pi)$ is the orbital frequency since $\dot{\Phi} = \omega$. The constants $t_c$ and $\Phi_c$ are usually called the coalescence time and phase, respectively. Therefore, the Fourier phase can be written as
\begin{align}
    \Psi(f) = 2\Phi_c - 2 \pi f t_c + 2\pi \int^{\infty}_{\tfrac{f}{2}} (f - 2F) \frac{dt}{dF} \, dF \,,
    \label{dephasing_def}
\end{align}
%where we used that $F = {f}/{2}$. 
We can further massage Eq.~\eqref{dephasing_def} to put it into a form more amenable to computation. Let us begin by writing this equation in terms of the energy of the system,
\begin{equation}
    \frac{dt}{dF} = \frac{dE}{dF} \mathcal{L}^{-1}\,,
\end{equation}
where we used the balance law $\dot{E} \equiv -\mathcal{L}$ to relate the rate of change of the orbital energy of the binary $\dot{E}$ to the gravitational-wave luminosity $\mathcal{L}$. With this in hand, the Fourier phase is given by
\begin{equation}
    \Psi(f) = 2\pi f t_c - 2\Phi_c -2\pi \int^{\frac{f}{2}}_{\infty} (2F-f) \mathcal{L}^{-1} \frac{dE}{dF} \, dF \,.
    \label{psi_GR_integral}
\end{equation}

We thus have that, in the stationary-phase approximation, the phase of the gravitational wave is calculated as an integral of the binding energy $E$ of the system and the luminosity $\mathcal{L}$ of the emitted gravitational wave. However, the binding energy and the luminosity may have other contributions apart from the ones from general relativity, such as modifications in $E$ and $\mathcal{L}$ due to perturbations in the metric, the presence of a third body, modified theories of gravity, etc. Let us parameterize these contributions as
\begin{align}
    \mathcal{L} &= \mathcal{L}_{M} + \delta \mathcal{L}\,, \\[7pt]
    E &= E_{M} + \delta E\,,
\end{align}
where $E_{M}$ and $\mathcal{L}_{M}$ are the binding energy of the orbit and luminosity of the gravitational wave respectively in general relativity and to as high a post-Newtonian order as one wishes to include. The quantities $\delta E$ and $\delta \mathcal{L}$ represent corrections to $E$ and ${\cal{L}}$ that are not from the binary. In this work, we focus on corrections that are only due to the presence of a vector-field dark-matter environment. As we explained earlier, the vector-field effect is conservative, because the perturbing force can be derived from the spatial gradient of a potential that is even under time-reversal. In particular, the vector-field environment that we consider does not induce radiative effects, since on the scales of interest to use, the vector-field is not a propagating wave that carries energy-momentum away from the binary. Therefore, the energy carried away in gravitational waves is the only source of energy loss. With this in mind, we conclude that $\delta \mathcal{L} = 0$ and $\delta E = V_A$, where $V_A$ is given in Eq.~(\ref{binding_energy_A}).

The vector field generates a dephasing in the waveform through a perturbation in the binding energy of the orbit. %This perturbation is given by a variation in the \textit{semi-major axis} of the orbit. %In particular, we will have a contribution $\delta E$ given by the change in the semi-major axis $\dot{a}$ due to the tangential force of the vector field on the orbit.
To calculate such a modification, we start by writing the phase as
\begin{equation}
    \Psi(f) = \Psi_M(f) + \Psi_A(f)\,,
\end{equation}
where $\Psi_M(f)$ is the Fourier gravitational-wave phase to as high a post-Newtonian order as one wishes to include, while $\Psi_A(f)$ is the dephasing correction generated by the vector-field environment. To find $\Psi_A(f)$, we insert the perturbed binding energy $\delta E =V_A$ of Eq.~(\ref{binding_energy_A}) into Eq.~\eqref{psi_GR_integral} to obtain
\begin{equation}
    \Psi_A(f) =  -2\pi \int^{\frac{f}{2}} (2F-f) \mathcal{L}_{\rm{M}}^{-1} \frac{d V_A}{dF} \, dF\,.
    \label{psi_perturbation}
\end{equation}
Observe that the evaluation of this integral requires that we convert between semi-major axis $b$ and frequency $F$, which we can do through Kepler's third law to leading post-Newtonian order (i.e.~$\omega^2 = M/b^3$), because the vector-field perturbing force does not correct this expression. 
%{\ny{Check you agree.}}\tc{I agree}

There is another simple way to calculate the dephasing which consists in calculating the perturbation in the orbital frequency of the binary as a function of time due to the presence of the vector field. That is, we split the variation of the frequency in a contribution from general relativity\footnote{At leading order in post newtonian perturbation theory, the variation of the frequency in general relativity is given by $\dot{F}_{M} = \tfrac{3}{2} F (\dot{b}/{b})_M$, with $(\dot{b}/{b})_M$ given in Eq. (\ref{eq:peters}).} plus the perturbation from the vector field, $\dot{F} = \dot{F}_{M} + \dot{F}_A$, with $\dot{F}_{M} \gg \dot{F}_A$. Then, we can calculate the dephasing as
\begin{equation}
    \Psi_A(f) =  - 2\pi \int^{\frac{f}{2}} (2F-f)  \frac{\dot{F}_A}{\dot{F}_M^2} dF\,.
    \label{psi_perturbation_bis}
\end{equation}
We can now use Kepler law to relate the variation in the frequency with a variation in the semi-major axes, which we calculated in Eq. (\ref{adotovera}), as $\dot{F}_A = -\tfrac{3}{2} F (\dot{b}/{b})_A$. In this way, we obtain an equivalent expression as the one given in Eq. (\ref{psi_perturbation}).
% The binding energy at zero order is given by the usual Kepler relation $E = - \mu M/2b$, since it is the Newtonian zero order solution of Eq.~(\ref{eq_newton_binary}). We thus have that the perturbation in the waveform due to the vector-field dark-matter environment is produced by a variation in the semi-major axis of the orbit $\dot{b}$.

With this in hand, the Fourier transform of the gravitational waves emitted by a quasi-circular inspiraling compact binary in a vector dark matter field environment, can be written as 
\begin{equation}
    \tilde{h}(f) = \tilde{h}_M(f) e^{i \Psi_A(f)}\,,
    \label{eq:full-h}
\end{equation}
in the inspiral phase, where $\tilde{h}_M(f)$ is the Fourier transform of the gravitational wave combination $h$ for a compact binary in isolation, in vacuum and in general relativity. One can in fact use the highest post-Newtonian order expression for the Fourier phase $\Psi_M(f)$ in $\tilde{h}_M(f)$ when evaluating $\tilde{h}(f)$ if one wishes. 

%------------------------------------------------------
\subsection{Calculation of the dephasing $\Psi_A(f)$} \label{sec:calculation_dephasing}

In order to solve the integral of Eq.~(\ref{psi_perturbation}), we need to calculate the variation of the binding energy with respect to the frequency. To do so, we follow the method of osculating orbits, and assume that the binding energy, to leading order, is given by the Newtonian expression, $E_M = -G\mu M/2b$. Then, the variations in the binding energy of the system are given by variations in the semi-major axes. Using the chain rule, we have that
\begin{equation}
    \frac{d E_A}{dF} = \left(\frac{d E_A}{dt}\right) \left(\frac{d t}{dF}\right) =  
    - E_M \left(\frac{dt}{dF}\right)_{M} \left(\frac{\dot{b}}{b}\right)_A\,,
\end{equation}
where $(\dot{b}/b)_A$ is given by Eq.~(\ref{adotoveravec}), while
\begin{equation}
    \left(\frac{dt}{dF}\right)_{M} = -\frac{dE_M}{dF} \frac{1}{{\cal{L}}_{M}} = - \frac{5 \pi}{48}\, \eta^{-1}M^2 (2\pi M F)^{-\frac{11}{3}}\,.
    \label{dtdFGR}
\end{equation}
Inserting the above expressions in the integral of Eq.~\eqref{psi_perturbation}, and changing variables to $x = (2\pi F \mathcal{M})^{\frac{1}{3}}$ (with $\mchirp = \eta^{3/5}M$ the chirp mass), we have
\begin{align}
    &\Psi_A(f) = \frac{25\pi G}{64}\sin(\alpha)^2 \mchirp^2 \rho_A \int_{\infty}^{x(f/2)} dx\, (\pi f\mchirp - x^3) \nonumber\\[5pt]
    &\times \frac{1}{x^{20}} \cos\left(\frac{5 \mchirp m_A}{128 x^8}+\gamma\right)\sin\left(\frac{5}{128 x^5}-2\varphi\right)\,,
    \label{psi_GR_integral_bis}
\end{align}
where we typically have $0.01 < x < 0.3$ for systems in the LISA band.

\begin{figure}[th]
    \centering
    \includegraphics[width=\linewidth]{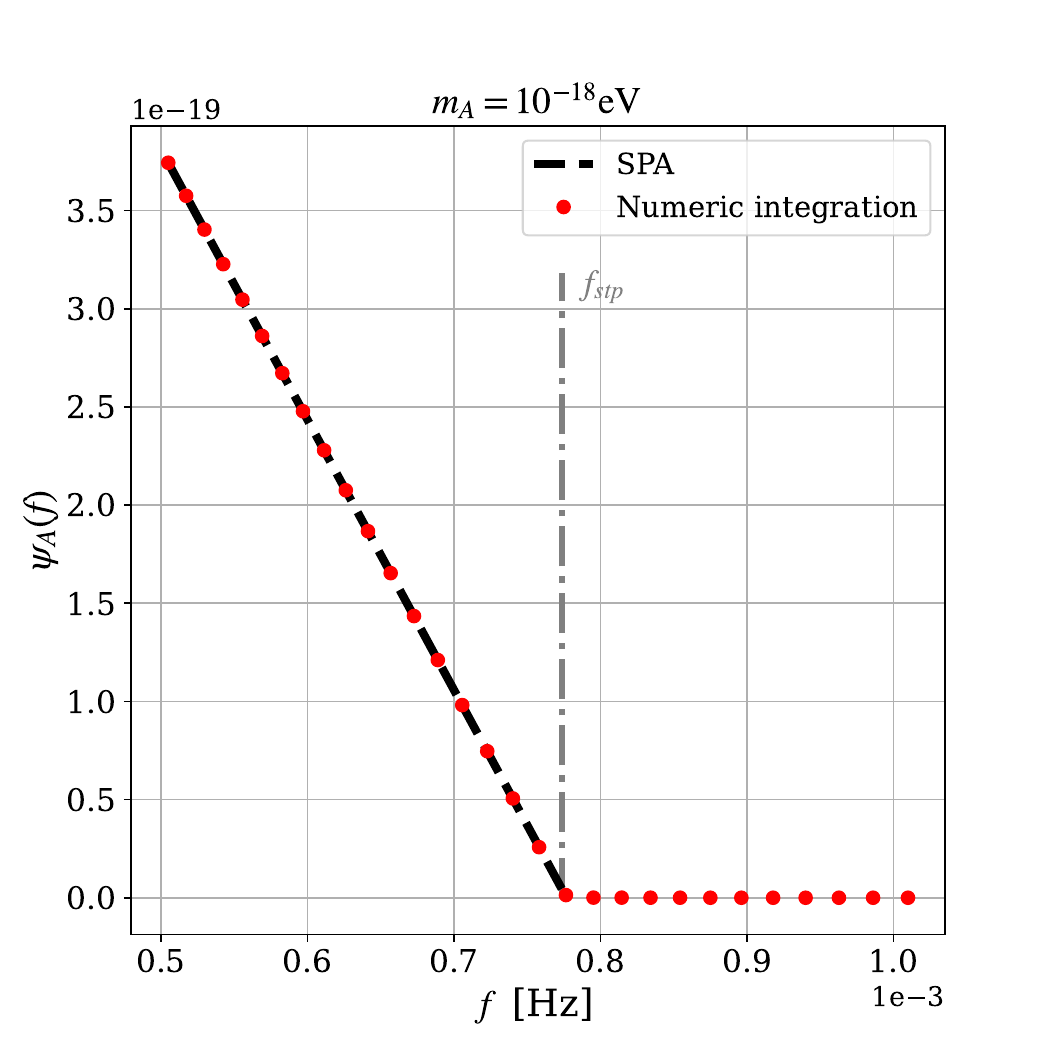}
    \caption{Dephasing of the gravitational wave calculated as in Eq.~(\ref{psi_GR_integral_bis}) under the stationary-phase approximation (black dashed line) and numerically integrated (red dots), for $m = 10^{-18}\rm{eV}$, $M=10^5 \rm{\rm{M}_{\odot}}$, $\eta=0.25$, $\varphi = 0$ and $\alpha = \pi/4$. The vertical dash-dot gray line corresponds to the stationary point $\pi f_{\rm{stp}} = \frac{8}{5} m_A$. For $f>f_{\rm{stp}}$ the integral averages to zero. Observe that the stationary-phase approximation is very accurate, with percentage errors $\ll 1\%$ in the range shown.}
    \label{fig:SPA_vs_Numeric}
\end{figure}

The above integral is highly oscillatory for the systems considered in this work, and thus, its solution is non-trivial. The oscillations have two characteristic time scales: one that is controlled by the chirp mass, and another that is dominated by the vector field mass. We thus solve the integral using the stationary-phase approximation again. We start by writing the integral as
\begin{equation}
    \Psi_A(f) = \int_{\infty}^{x(f/2)} dx \,\mathcal{A}(x) e^{i \phi(x)}\,,
\end{equation}
where $\mathcal{A}(x)$ is a slowly varying amplitude and 
\begin{equation}
    \phi(x) = \frac{5}{128 x^8}\left(x^3-\mchirp m_A\right) - 2 \gamma \,.
\end{equation}
We look for stationary points where $\phi^{\prime}(x_{\rm{stp}}) = 0$, which yields $\pi f_{\rm{stp}} = 8 m_A/5$. Observe that the stationary point depends on the vector field mass $m_A$, but it is independent of the total mass of the binary $M$. Since we are integrating from $F=+\infty$, the integral is approximately zero until the stationary point $F \sim f_{\rm{stp}}$ is reached, so $\Psi_A(f) = 0$ for $F \gg f_{\rm{stp}}$. Once the stationary point is crossed, the integral has a contribution from the neighborhood of $f_{\rm{stp}}$ given by
\begin{align}
    \Psi_A(f) = &\sqrt{\frac{5}{6}}\frac{15625 \,\pi^{5/2} G\,\rho_A}{524288 \mchirp^{5/2} m_A^{11/2}} (f_{\rm{stp}} - f) \times \label{delta_psi_A}\\[3pt]
    &\times \sin(\alpha)^2 \sin(\xi - \frac{\pi}{4} + 2 \varphi)\,,\nonumber
\end{align}
where we have defined $\xi = ({75}/{32768}) 5^{2/3}(\mchirp m_A)^{-5/3}- \gamma$. For $F \ll f_{\rm{stp}}$, the integrand averages to zero again so the only contribution to the dephasing comes from the neighborhood of $F\sim f_{\rm{stp}}$, given by Eq.~(\ref{delta_psi_A}).
Figure~\ref{fig:SPA_vs_Numeric} shows the dephasing for an equal-mass binary with total mass $10^5 \rm{M}_\odot$, and a vector field of mass $m_A = 10^{-18}\rm{eV}$. We calculate the dephasing under the stationary-phase approximation, and we also do so numerically, by explicitly integrating Eq.~(\ref{psi_GR_integral_bis}). Observe that our stationary-phase approximation to the solution of the integral accurately matches the numerical solution. The percentage error between the stationary-phase approximation and the numeric approach is $\ll 1\%$. 

In the life cycle of binaries, the orbital (and gravitational wave) frequency chirps from small values, associated with the inspiral, to large values associated with the merger. The initial minimum value $f_{\rm{min}}$ is that at which we begin to observe the gravitational wave signal; we here take $f_{\rm{min}}$ to be either the lowest observable frequency in the LISA band, roughly $10^{-5}$ Hz, or the frequency 4 years before reaching merger. The maximum value $f_{\rm{max}}$ is that associated with the merger, which we can think of as roughly the innermost stable circular orbit (ISCO) frequency, $f_{\rm ISCO}$, for a test particle in a Schwarzschild black hole. Then, the solution given in Eq.~(\ref{delta_psi_A}) is valid for vector-field dark matter masses that are contained in the LISA band, namely $10^{-19}\rm{eV} \lesssim m_A \lesssim 10^{-15}\rm{eV}$, and for systems where $f_{\rm{ISCO}} > f_{\rm{stp}}$. For systems such that $f_{\rm{stp}} < f_{\rm{min}}$ or $f_{\rm{ISCO}} < f_{\rm{stp}}$, the effect of the vector-field dark matter on the gravitational wave averages to zero. 

%%%%%%%%%%%%%%%%%%%%%%%%%%%%%%%%%%%%%%%%%%%%%%%%%%%%%%%%%%%%%%%%%%%%%%%%%%%%%%%%
\section{Fisher analysis} \label{sec:Fisher_analysis}

In this section, we present a Fisher analysis to compute the parameter space for which the dark-matter effect calculated in this manuscript can be detected in a gravitational-wave observation. We start by defining the inner product between two waveforms $h$ and $g$ via
\begin{equation}
    \langle h | g \rangle = 2 \int_{f_{\rm{min}}}^{f_{\rm{max}}}  \frac{\tilde{h}^* \tilde{g} + \tilde{h} \tilde{g}^*}{S_n} df\,,
\end{equation}
where $S_n(f)$ is the \textit{power noise spectrum} of the instrument, the overhead tilde stands for the Fourier transform, and the superscript star for complex conjugation. In this work, we focus on binary systems that can be detected in the LISA band, so we perform our analyses using an approximation to the LISA power noise spectrum \cite{Robson:2018ifk}. 

With the inner product in hand, one can now define the \textit{signal-to-noise ratio} (S/N) for a given signal $h$ as
\begin{equation}
    \rho[h] = \langle h | h\rangle^{1/2}\,.
\end{equation}
In general, the signal $h$ depends on a set of parameters $\{{\theta}_i\}$ which are estimated by \textit{matched filtering}. In the limit of $\rho[h] \gg 1$, and for Gaussian and stationary noise, one can show that the $1\sigma$ uncertainty in the estimation of parameters is given by
\begin{equation}
    \Delta \theta^a \equiv \sqrt{\langle (\theta^a - \langle \theta^a\rangle)^2\rangle} = \sqrt{\Sigma^{aa}}\,,
    \label{eq:variance}
\end{equation}
where $\Sigma^{aa}$ is the $(a,a$) diagonal element of the inverse of the covariance matrix (and summation is not implied here). The covariance matrix, also known as the \textit{Fisher information matrix} $\Gamma_{ab}$, is defined as
\begin{equation}
    \Gamma_{ab} = \left\langle \frac{\partial h}{\partial \theta_a} \bigg| \frac{\partial h}{\partial \theta_b} \right\rangle \,.
    \label{inner_product_def}
\end{equation}

We wish to perform a Fisher analysis to estimate the accuracy to which we could measure dark-matter parameters contained in the dephasing of Eq.~\eqref{delta_psi_A} with a future LISA gravitational-wave observation. To do so, it is convenient to parameterize the dephasing of the gravitational wave in terms of two parameters, as
\begin{equation}
    \Psi_A(f) = \beta \left(1 - \frac{f}{f_{\rm{stp}}}\right)\theta(f_{\rm{stp}} - f)\,,
\end{equation}
 where $\pi f_{\rm{stp}} = \frac{8}{5} m_A$ and $\beta$ is given by
\begin{align}
    \beta &\sim \sqrt{\frac{5}{6}}\frac{625 \,\pi^{3/2} G\,\rho_A}{65536 \mchirp^{5/2} m_A^{9/2}} \sin(\alpha)^2 \label{eq_beta}\\[4pt]
    &\sim 3 \times10^{-12}\frac{\rho_A}{0.3\,\frac{\mathrm{GeV}}{\mathrm{cm^3}}}\left(\frac{10^{4}\mathrm{\rm{M}_{\odot}}}{M}\right)^{5/2}\left(\frac{10^{-19}\mathrm{eV}}{m_A}\right)^{9/2}\,.\nonumber  
\end{align}
We then add these two parameters to the rest of the parameters contained in the vacuum part of the gravitational wave signal $\tilde{h}_{M}$ in Eq.~\eqref{eq:full-h}, which in this paper we will model through a TaylorF2 model~\cite{Damour:2000zb} valid only up to the ISCO, namely
\begin{align}
    \Psi_M &= 2 \pi  f t_c-\Phi _c -\frac{\pi }{4} + \frac{3}{128} U^{-5/3} \\[4pt]
    &-\frac{3 \pi}{8}  \eta^{-3/5} U^{-2/3}  + \frac{5}{96}\eta ^{-2/5} \left(\frac{743}{336} + \frac{11 \eta }{4}\right)U^{-1} \nonumber\,,
\end{align}
with $U = \pi \mchirp f$.
Equation~\eqref{eq:full-h} then gives us our dark-matter enhanced waveform model. 

The Fisher estimate of the accuracy to which dark-matter parameters can be measured with a future gravitational wave observation can be carried out in various ways. In this paper, we consider the hypothetical scenario that the true gravitational wave signal did \textit{not} contain any dark-matter effects (e.g.~because there was no dark matter near this particular binary), and then we ask how well we can ascertain that the dark-matter density $\rho_A$ (or the dephasing parameter $\beta$) is consistent with zero. In a Bayesian language, we expect that a Bayesian parameter estimation study of such a hypothetical signal with our model would yield approximately-Gaussian marginalized posteriors centered at $\beta=0$, and we wish to determine the width of this distribution. The $1\sigma$ width is then given by Eq.~\eqref{eq:variance}, with the Fisher matrix evaluated for the $a = \beta$ parameter, and setting $\beta=0$ after taking the $\beta$-derivatives. 

From these considerations, it is clear that we first need to calculate the following derivatives (beside the derivatives with respect to the usual vacuum parameters that characterize $\tilde{h}_M$) to compute the Fisher matrix:
\begin{align}
    \frac{\partial \tilde{h}(f)}{\partial\beta} &= i\left(1 - \frac{f}{f_{\rm{stp}}}\right)\theta(f - f_{\rm{stp}})\, \tilde{h}(f)\,,\\[5pt]
    \frac{\partial \tilde{h}(f)}{\partial f_{\rm{stp}}} &= i\beta\bigg[\frac{f}{f_{\rm{stp}}^2} \theta(f - f_{\rm{stp}}) + \left(1 - \frac{f}{f_{\rm{stp}}}\right) \\[3pt]
    &\qquad\times\delta(f-f_{\rm{stp}})\bigg] \,\tilde{h}(f)\,. \nonumber
\end{align}
Observe that the derivative with respect to $f_{\rm{stp}}$ does not contribute to the Fisher matrix because we perform the analysis for the fiducial value $\beta=0$, so every matrix element containing such derivative vanishes. 
Therefore, we drop $f_{stp}$ from the parameters list to perform the Fisher analysis. 

For systems with $f_{\rm{stp}} < f_{\rm{min}}$, the effect of the vector-field dark matter on the orbit averages to zero; this is the region where the step function always vanishes. On the other hand, for binaries with $f_{\rm{ISCO}} < f_{\rm{stp}}$, the dephasing is degenerate with $\phi_c$ and $t_c$, and thus $\beta$ also cannot be measured. However, for systems whose frequency chirps \textit{through} the stationary frequency $f_{\rm{stp}}$ in the period of observation, the degeneracy is broken as a consequence of the step function turning on during the observation time. We thus consider systems for which, in the period of observation, the frequency takes the value $f_{\rm{stp}}$ at some point in the chirp of the binary. 

\begin{figure}[th]
    \centering
    \includegraphics[width=\linewidth]{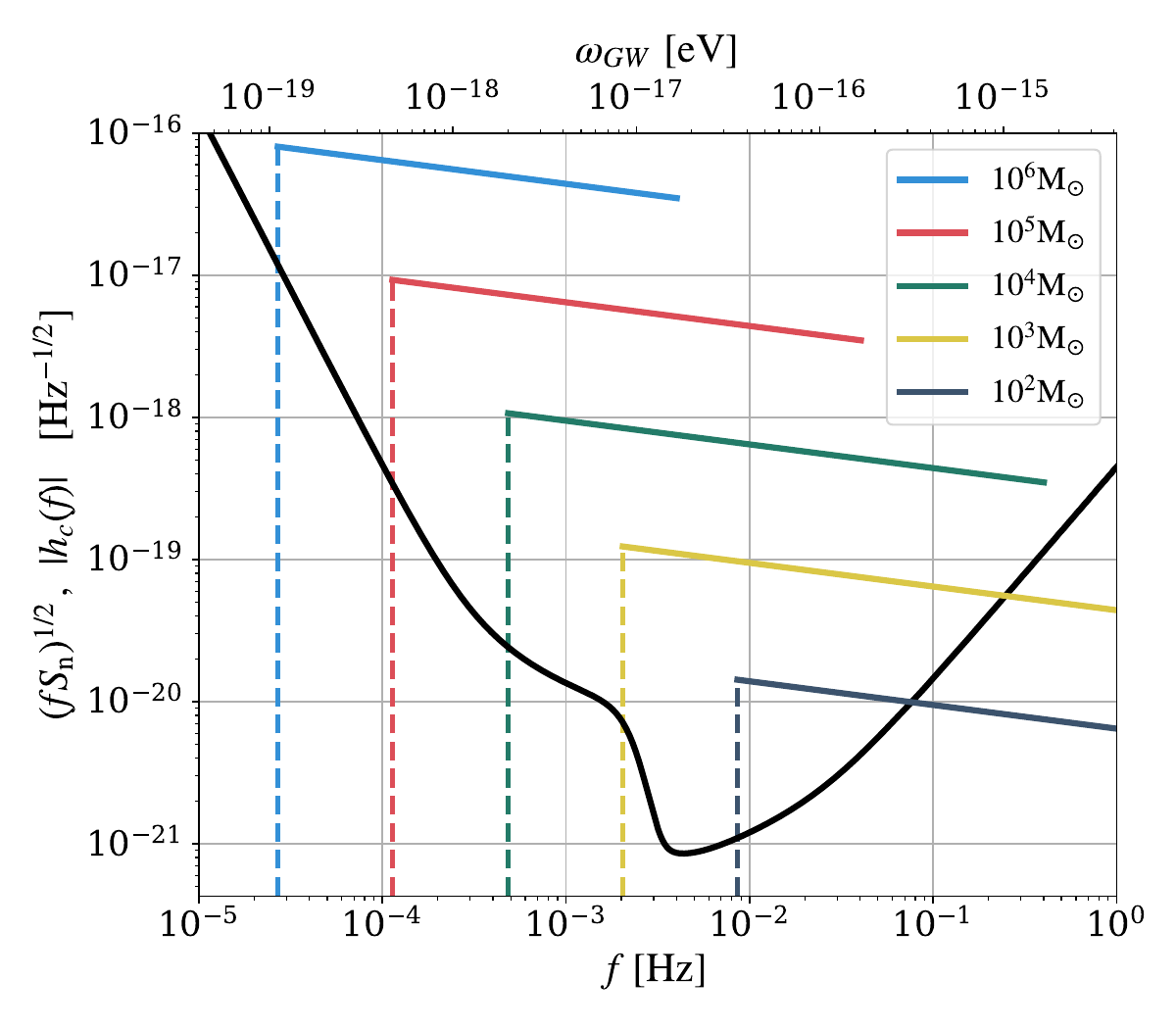}
    \caption{LISA's sensitivity strain (black curve). The different colors indicate the gravitational wave characteristic strain for each binary mass during the inspiral, defined as $|h_c (f)|= 2 f |\mathcal{A}(f)|$. We consider symmetric-mass binaries at $D_L = 1 \rm{Gpc}$. The mass of each member of the binary is the one indicated in the label. The dashed lines indicate $f_{\rm{min}}$ for each system, calculated as the frequency four years before the ISCO is reached (corresponding to LISA's observation period). On the top axes we show $\omega_{GW} \equiv 2 \pi f$.}
    \label{fig:Sn_mass}
\end{figure}

Figure~\ref{fig:Sn_mass} shows the gravitational-wave characteristic strain for different binary masses as a function of the gravitational-wave frequency. The black solid curve is the characteristic LISA sensitivity strain~\cite{Robson:2018ifk}. For each binary mass, we show the frequency range each binary sweeps through. The initial frequency $f_{\rm{min}}$ is calculated as the frequency the binary had four year before ISCO, corresponding to a large observation period with LISA. Observe that each frequency range also corresponds to the mass range of the vector field that could be tested with a gravitational-wave observation. This is because, as mentioned before, for each binary evolution, the orbital frequency resonates with the vector field mass at some point in the evolution of the binary, activating the step function and thus breaking the degeneracy. In particular, we see that for systems that could be probed by LISA, the mass range for the vector field that could be tested in a detection is $m_A \sim (10^{-19}, 10^{-15})\rm{eV}$. 

\begin{figure}[th]
    \centering
    \includegraphics[width=\linewidth]{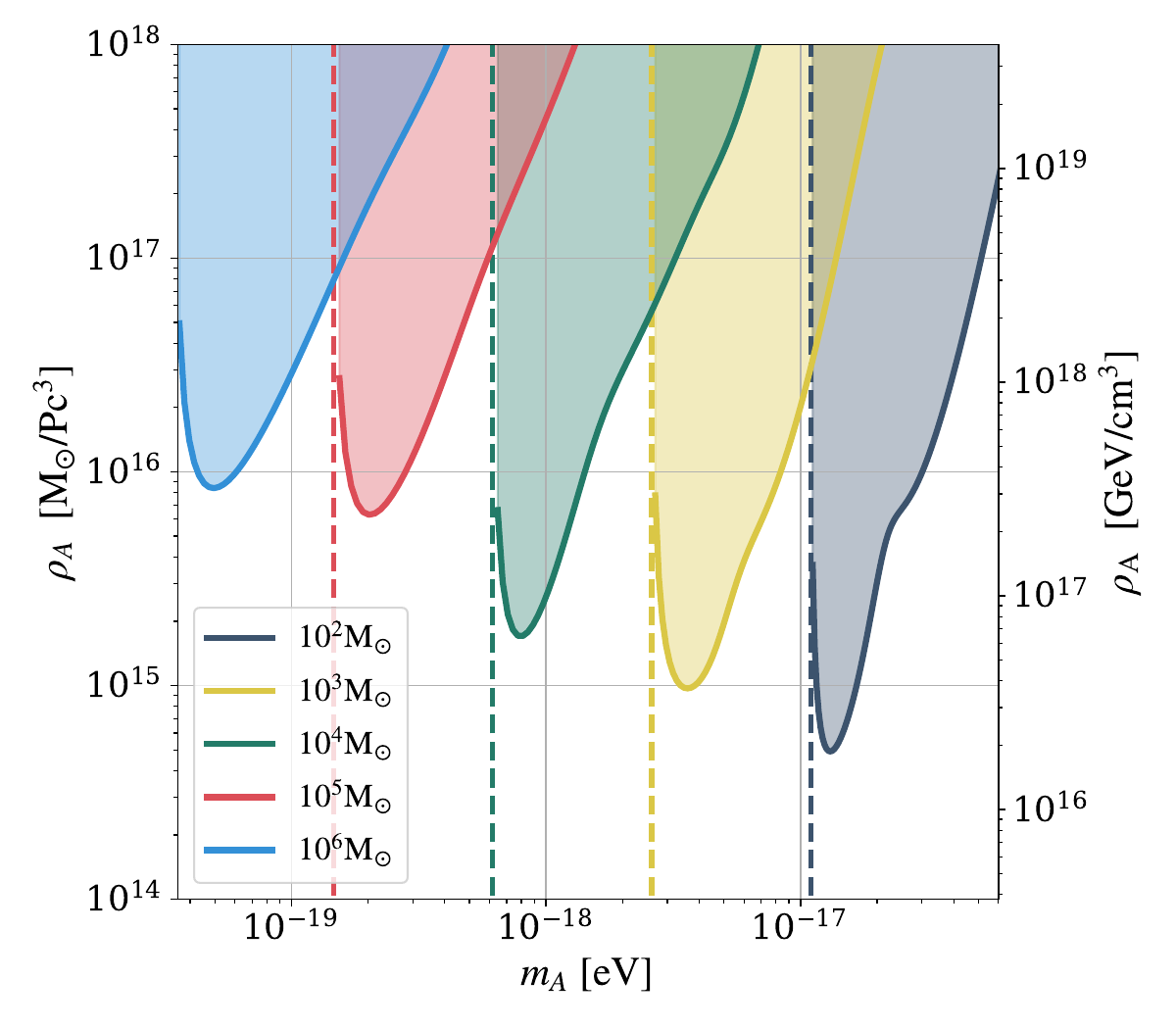}
    \caption{$1\sigma$ accuracy (estimated with a Fisher analysis) to which $\rho_A$ can be measured, given a value of the vector mass in eV, for a binary with various total masses (shown with different colors). The vertical dash-dot lines indicates $f_{\min}$, corresponding to the frequency of the system four year before the ISCO. We assume $\eta = 0.25$ for the analysis. If a binary system is immersed in a dark-matter vector environment with a density larger than these values (shaded regions), then the gravitational waves emitted carry a dark-matter signature that is detectable with LISA for a 4-year observation.}
    \label{fig:fisher}
\end{figure}

\begin{figure}[th]
    \centering
    \includegraphics[width=\linewidth]{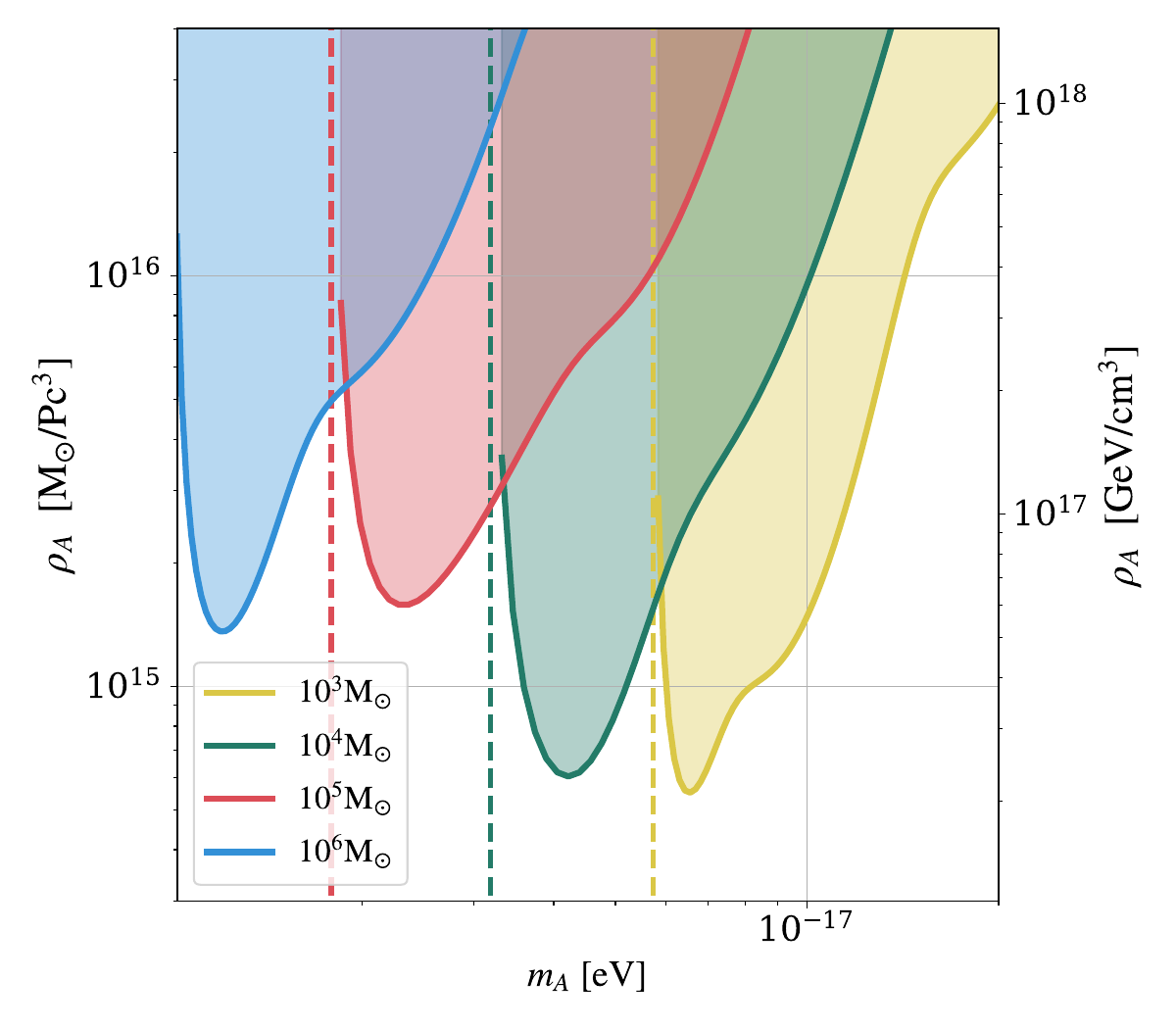}
    \caption{$1\sigma$ accuracy (estimated with a Fisher analysis) to which $\rho_A$ can be measured, given a value of the vector mass in eV, for a binary with various total masses (shown with different colors). The vertical dash-dot lines indicates $f_{\min}$, corresponding to the frequency of the system four year before the ISCO. For this analysis, we assume binary systems with assymetric masses, where the heavier mass is given for each color, and the lightest mass is always $10^2 \rm{M}_{\odot}$. If a binary system is immersed in a dark-matter vector environment with a density larger than these values (shaded regions), then the gravitational waves emitted carry a dark-matter signature that is detectable with LISA for a 4-year observation. We can see that the effect is greater than in the symmetric case (Fig. \ref{fig:fisher}).}
    \label{fig:fisher_assymetric}
\end{figure}

Figure~\ref{fig:fisher} shows the $1\sigma$ accuracy (estimated with a Fisher analysis) to which $\rho_A$ can be measured, as a function of the vector field mass for 5 different future LISA observations of equal mass binaries with different total mass. The shaded regions corresponds to values of the dark-matter density for which the vector-field dark-matter effects we considered in this paper would leave a measurable (with LISA) impact on the gravitational waves emitted. Although the fiducial value for the dark-matter density in our Solar System is $\rho_{\rm{DM}} \sim 0.3 \rm{GeV}/\rm{cm}^3$, this density could be significantly larger near supermassive black holes.   We thus see that LISA opens the window to probing a new regime of vector-dark-matter densities with gravitational waves. On Figure \ref{fig:fisher_assymetric} we show the same analysis as for Fig. \ref{fig:fisher}, but considering assymetric systems. In particular, we consider systems where the lighter BH in the binary has a mass of $10^2 \rm{M}_{\odot}$, while the heavier companion's mass is given by the label of each curve. We can see that, for assymetric masses, the dark matter effect is enhanced with respect to symmetric binaries, as was expected from Eq. (\ref{eq:ratio_bdots}).

%%%%%%%%%%%%%%%%%%%%%%%%%%%%%%%%%%%%%%%%%%%%%%%%%%%%%%%%%%%%%%%%%%%%%%%%%%%%%%%%
\section{Conclusions}

In this work, we studied the dephasing in the gravitational waves emitted by compact binaries due to an ultralight vector dark-matter field environment. Ultralight fields produce oscillating gravitational potentials that are approximately homogeneous at the characteristic scales of compact binary systems. These oscillations perturb the orbit, producing a variation in the binding energy of the system. We show that such perturbations generate a dephasing in the gravitational waves emitted by the binary and calculate explicitly the Fourier phase of the signal. We find that the dark-matter correction to the Fourier phase is proportional to the dark-matter energy density in the local patch, $\rho_A$. We also show that the dark-matter effect can be important when the frequency of the gravitational wave resonates with the vector field mass at some point in the evolution of the binary. 

We performed a Fisher analysis to estimate the accuracy to which we may be able to constrain the dark-matter density $\rho_A$ with a future LISA observation. We found that 4-year LISA observations should allow us to measure dark-matter densities larger than $10^{16} \; {\rm{GeV}}/{\rm{cm}}^3$ for vector masses in the range $(10^{-19},10^{-16}) \; {\rm{eV}}$. We also find that these measurements are best for asymmetric binaries with low total mass. Interestingly, some of these are systems whose merger could potentially also be observed by LIGO, after LISA observes the early inspiral~\cite{Sesana:2016ljz}. Such multi-wavelength gravitational-wave events seem ideal to probe non-vacuum or beyond-Einstein effects in gravitational wave signals~\cite{Barausse:2016eii,Perkins:2020tra}.
Therefore, future LISA observations of the gravitational waves emitted in the quasi-circular inspiral of massive black holes may detect or yield interesting constraints on the density of vector-type ultralight dark matter surrounding binary black holes systems.   

Future work could focus on a complete study of the perturbations in the gravitational waveform due to the presence of a vector field environment. Such a study should include perturbations in the waveform resulting from the variation of other orbital parameters, in addition to the variation of the semi-major axes considered here. In particular, future work could focus on the variation of the inclination of the orbital plane, which has an impact on the amplitude of the gravitational wave. Also, the formalism presented in this manuscript can be generalized for eccentric orbits \cite{Yunes:2009yz}, since the vector field also has an impact on the evolution of the eccentricity of the orbits.

%%%%%%%%%%%%%%%%%%%%%%%%%%%%%%%%%%%%%%%%%%%%%%%%%%%%%%%
\acknowledgements
We thank Esteban Calzetta for useful discussions. N.~Y.~ acknowledges support from the Simons Foundation through Award No. 896696, the NSF through Grant No. PHY-2207650, and NASA through Grant No.~80NSSC22K0806. T.~F.~C.~ and D.~L.~N.~ acknowledges support from CONICET and UBA.
%{\ny{Insert others for Diana?}}

%%%%%%%%%%%%%%%%%%%%%%%%%%%%%%%%%%%%%%%%%%%%%%%%%%%%%%%%%%%%%%%%%%%%%%%%%%%%%%%%
\appendix

%%%%%%%%%%%%%%%%%%%%%%%%%%%%%%%%%%%%%%%%%%%%%%%%%%%%%%%%%%%%%%%%%%%%%%%%%%%%%%%%
\section{Binary systems in ultralight dark matter halos} \label{appendix:backreaction}

In this appendix, we discuss the validity of approximating the dark matter energy profile as homogeneous on the extension of the orbit.  Numerical solutions show  that dark matter halos in ultralight dark matter models can be described by solitonic cores surrounded by  Navarro-Frenk-White profiles. The radial density profile of the core can be modeled as \cite{Schive:2014hza, Schive:2014dra, Schwabe:2016rze}
\begin{equation}
    \rho_{c} \sim \rho_{c, 0} [1 + 0.091 (r/r_c)]^{-8}\,,
    \label{core_energy_density}
\end{equation}
with
\begin{equation}
    \rho_{c, 0} \sim 1.9 \times 10^{-10} \left(\frac{10^{-18}\rm{eV}}{m_{A}}\right)^2 \left(\frac{\rm{kpc}}{r_c}\right)^4 \frac{\rm{M}_{\odot}}{\rm{pc}^3}\,,
\end{equation}
where $r_c$ is the core radius, defined as the radius at which the density drops to half of its peak value. The mass of the core is defined as the mass enclosed by $r_c$. %
For ultralight scalar field models, the core radius can be related to dark-matter field mass and the corresponding total halo masses. At redshift $z=0$, $r_c$ can be estimated as\footnote{See for instance Eq. (7) in Ref. \cite{Schive:2014hza}.}
\begin{equation}
    r_c = 10^{11} \left(\frac{10^{-18}\rm{eV}}{m_A}\right) \left(\frac{M_{\rm{halo}}}{10^{12} \rm{M}_{\odot}} \right)^{-1/3} \rm{km}\,.
  \label{core_halo_relation}\end{equation}
If we suppose that each halo hosts a galaxy with a supermassive black  at its center \cite{Kormendy:1995er,Ferrarese:2000se}, we can further relate the mass of the black hole with the mass of the  dark matter halo  as \cite{Ferrarese:2002ct, 10.1111/j.1745-3933.2010.00832.x, Bandara_2009}
\begin{equation}
    M \sim 10^7 \rm{M}_{\odot} \left(\frac{M_{\rm{halo}}}{10^{12}\rm{M}_{\odot}}\right)^{\kappa},
    \label{halo_mass}
\end{equation}
where $\kappa \sim 1.6$. 

In order to explain our set up, we consider the case of supermassive and low/intermediate massive binaries separately. In the case of supermassive black hole binaries, it is reasonable to assume that, due to processes such as dynamical friction and mass segregation, the black holes are most probably located near the center of their host galaxy. Moreover, these galaxies themselves are typically located near the core centers of their respective halos \cite{Kravtsov:2003sg, Berlind:2002rn}, where the energy density is approximately constant (see Eq. \ref{core_energy_density}). Then, if the size of the orbit $b$ is much smaller than the size of the core $r_c$, we can assume that the energy profile throughout the binary is approximately homogeneous. To check the validity of this assumption, we consider a supermassive binary of $M=10^6\rm{M}_{\odot}$, which corresponds to the largest system LISA can probe. Using Eq. (\ref{halo_mass}) for this system we obtain a halo mass of approximately $M_{\rm{halo}} = 10^{11} \rm{M}_{\odot}$; and from Eq. (\ref{core_halo_relation}),  for a field environment of mass $m_A = 10^{-18}\rm{eV}$, we obtain  a corresponding central core of radius $r_c = 2.5\times 10^{11}\rm{km}$. We thus see that the core size is much larger than the orbits considered in this work (see Eq. \ref{typical_radius}), so we can assume an approximately homogeneous energy density profile in the extension of the orbit. Also, we see that this estimate matches the one from the discussion in Sec. \ref{sec:impact_vector-field dark matter_orbit} regarding the de Broglie wavelength given in Eq. (\ref{lambda_dB}), since we used the velocity dispersion of the Milky Way, which hosts a supermassive black hole at its center (Sag.~${\rm{A}}^\star$) of $\sim4\times10^6 \rm{M}_{\odot}$, and has a halo mass of $\sim7\times10^{11}\rm{M}_{\odot}$.

For supermassive binaries, there is also non-negligible backreaction of the binary on the dark-matter field environment. Numerical simulations of inspiralling supermassive binaries in ultralight dark-matter field environments show that this backreaction generates a quasi-stationary energy-density profile around the binary, which is approximately homogeneous throughout the extension of the orbit, and is peaked in the vicinity of the black holes (see for instance Refs. \cite{Bamber:2022pbs, Aurrekoetxea:2023jwk,Boey:2025qbo}). 
This energy profile can be understood as a local enhancement of the host soliton density profile, produced by the accretion of dark matter by the black holes.  The effect is maximal when the orbital radius $r_{12}$ is of order of the Comptom wavelenth  $\lambda_c$ of the field.   In our case, the orbital radius of the systems satisfies   $r_{12}\ll \lambda_c$.  We here ignore this effect, so the values used for the density of the dark matter field in the orbit are in general underestimated.  A larger density would only enhance the effects considered here, making our estimates necessarily conservative.

Finally, we consider the case of intermediate- and low-mass black hole binaries. These systems are much smaller in spatial extent than the supermassive ones, since the radius of the orbit scales as the total mass of the binary. Even if they are not at the center of the core, it is most probable that the orbit is still completely embedded in it, or in an interference granule of characteristic size given by $\lambda_{dB}$. This means that they can be considered to be inside a region where the vector field is assumed to be homogeneous.
Furthermore, because of its low masses, the backreaction of the binary over the halo can be neglected. Then, we can assume that the vector field is homogeneous through the orbit of intermediate- and low-mass binaries, as done throughout this work.

\bibliographystyle{ieeetr} % We choose the "plain" reference style
\bibliography{references}

\end{document}